\documentclass[12pt,preprint]{aastex}

\newcommand{\degg}{\hbox{$^\circ$}}

\newcommand{\arcs}{\hbox{$^{\prime\prime}$}}

\newcommand{\et}{et al.\ }

\newcommand{\xmm}{{\it XMM-Newton}}

\newcommand{\chandra}{{\it Chandra}}
\newcommand{\sax}{{\it BeppoSAX}}
\newcommand{\ginga}{{\it Ginga}}

\newcommand{\asca}{{\it ASCA}}

\newcommand{\ls}
{\mathrel{\hbox{\rlap{\hbox{\lower4pt\hbox{$\sim$}}}\hbox{$<$}}}}
\newcommand{\gs}
{\mathrel{\hbox{\rlap{\hbox{\lower4pt\hbox{$\sim$}}}\hbox{$>$}}}}

\received{2003 September 10}
\begin{document}

\title{The XMM-Newton Iron Line Profile of NGC 3783}
\shorttitle{The Iron Line Profile of NGC 3783}
\shortauthors{Reeves \et}
\author{ J.N. Reeves\altaffilmark{1,2}, K. Nandra\altaffilmark{3}, 
I.M. George\altaffilmark{1,4},
K.A. Pounds\altaffilmark{5},
T.J. Turner\altaffilmark{1,4}, T. Yaqoob\altaffilmark{1,6}}
\email{jnr@milkyway.gsfc.nasa.gov}

\altaffiltext{1}{Laboratory for High Energy Astrophysics, Code 662, 
NASA Goddard Space Flight Center, Greenbelt Road, Greenbelt, MD 20771, USA.}

\altaffiltext{2}{Universities Space Research Association}

\altaffiltext{3}{Astrophysics Group, Imperial College, 
Blackett Laboratory, Prince Consort Road, London SW7 2BW, UK.}

\altaffiltext{4}{Joint Center for Astrophysics, 
University of Maryland, Baltimore County, 1000 Hilltop Circle, Baltimore, 
MD 21250, USA.}

\altaffiltext{5}{X-ray and Observational Astronomy Group; University of
Leicester; Leicester LE1 7RH; UK.}

\altaffiltext{6}{Department of Physics and Astronomy, 
Johns Hopkins University, Baltimore, MD 21218, USA.}

\begin{abstract}

We report on observations of the iron K line in the nearby Seyfert 1 
galaxy, NGC 3783, obtained in a long, 2 orbit ($\sim240$~ks) 
\xmm\ observation. The line profile obtained exhibits two strong narrow peaks 
at 6.4 keV and at 7.0 keV, with measured line equivalent 
widths of 120 and 35 eV respectively. The 6.4 keV emission 
is the K$\alpha$ line from near neutral Fe, whilst the 7.0 keV 
feature probably originates 
from a blend of the neutral Fe K$\beta$ line and the Hydrogen-like
line of Fe at 6.97 keV. The relatively 
narrow velocity width of the K$\alpha$ line 
($\lesssim5000$~km~s$^{-1}$), its lack of response to the continuum emission 
on short timescales and the 
detection of a neutral Compton reflection component
are all consistent with a distant origin in Compton-thick matter 
such as the putative molecular torus. A strong 
absorption line from highly ionized iron (at 6.67 keV) is detected 
in the time-averaged iron line profile, whilst 
the depth of the feature appears to vary with time, being strongest when the 
continuum flux is higher. 
The iron absorption line probably arises from the highest 
ionization component of the known warm absorber in NGC 3783, 
with an ionization of log~$\xi\sim3$ and column density of 
$N_{\rm H}\sim5\times10^{22}$~cm$^{-2}$ 
and may originate from within 0.1~pc of the nucleus. 
A weak red-wing to the iron K line 
profile is also detected below 6.4 keV. However when the effect of the 
highly ionized warm absorber on the underlying continuum is 
taken into account, the requirement for 
a relativistic iron line component from the inner disk is reduced.

\end{abstract}


\section{Introduction}

NGC 3783 is a bright (V=13~mag), nearby (z=0.00973), 
Seyfert 1 galaxy, which was first detected in X-rays 
in the {\it Ariel-V} all sky survey \citep{McHardy81}
and subsequently in the high Galactic latitude survey
conducted by HEAO-1 \citep{Picc82}. Since these early X-ray detections, 
there have been many observations of NGC 3783 in the X-ray band.
A ROSAT observation of 
NGC~3783 showed evidence for a ionized absorber in the soft 
X-ray spectrum \citep{JTurner93}, which was confirmed during 
\asca\ observations \citep{George95, George98}. Subsequent  
high resolution grating observations of NGC~3783 with \chandra\ and \xmm\ 
\citep{Kaspi00, Kaspi01, Kaspi02, Blustin02, Behar03}
have revealed the soft X-ray absorber with unprecedented
accuracy and resolution. 
Indeed the recent 900 ks observation obtained with the Chandra 
High Energy Transmission Grating Spectrometer (HETGS)
probably represents the best soft 
X-ray spectrum (in terms of the combination of spectral resolution and 
signal to noise) obtained on any AGN to date, with the spectrum showing 
numerous absorption lines from all of the abundant elements between 
C and Fe, over a wide range of ionization states 
\citep{Kaspi02, Krongold03, Netzer03}.

Our primary aim in this paper is to study in detail the iron K line 
profile of NGC 3783. The iron K$\alpha$ emission line diagnostic in AGN 
first became important during the \ginga\ era, showing
that the 6.4 keV iron K$\alpha$ emission line and associated Compton 
reflection hump above 7 keV was common amongst Seyfert 
galaxies \citep{Pounds1990, Nandra94}. 
The higher (CCD) 
resolution spectra available with the \asca\ satellite appeared to 
indicate that the line profiles were broad and asymmetrically skewed (to 
lower energies), which was interpreted as evidence 
that the majority of the line emission 
originated from the inner accretion disk around the massive black hole 
\citep{Tanaka95, Nandra97, Reynolds97}. 
Indeed observations of 
NGC~3783 with \asca\ also appeared to show a broad, relativistic iron 
line profile \citep{Nandra97, George98}, whilst the presence of 
the higher energy Compton reflection hump in NGC 3783 
has been confirmed in a \sax\ observation \citep{DeRosa02}.

The picture now emerging from the study of the iron K line with 
\xmm\ and \chandra\ appears to be much more complex. The presence of 
a narrower 6.4 keV iron emission component, from more distant matter 
(e.g. the outer disk, BLR or the molecular 
torus) appears to be commonplace in many type I AGN including 
NGC 3783, e.g. Mrk 205 \citep{R01}; NGC 5548 \citep{Yaqoob01}; 
NGC 5506 \citep{Matt01}; Mrk 509 \citep{Pounds02}; 
NGC 3516 \citep{JTurner02}; NGC 4151 \citep{Schurch03} and many other objects.
In contrast the broad, relativistic component of the 
iron line profile 
appears to be much weaker than anticipated \citep{PR02, R03b} 
and in some cases may be absent altogether, 
e.g. NGC 5548 \citep{Pounds03a}; NGC 4151 \citep{Schurch03}. 
Observations have also revealed that highly ionized emission components 
(from He or H-like iron) 
may also be present in some, typically higher luminosity AGN, e.g. 
PG 1116+215 \citep{Nandra96}; Mrk 205 \citep{R01}; NGC 5506 \citep{Matt01}; 
Mrk 509 \citep{Pounds02}; Mrk 766 \citep{Pounds03b}; 
NGC 7314 \citep{Yaqoob03}. 
The situation seems even further complicated,
because of the presence of ionized iron K-shell 
absorption edges and/or lines in some AGN 
\citep{Nandra99, Chartas02, Chartas03, Pounds03c, R03}, 
which may be associated with high velocity 
outflows. Furthermore transient, narrow, redshifted Fe line features 
have been observed in some AGN, e.g. NGC 3516 \citep{JTurner02}; Mrk 766, 
\citep{JTurner03}.

Here we present a study of the iron K-shell line 
from a long (240 ks) observation of NGC 3783 conducted by \xmm\ in 
December 2001. The results from a much shorter (40 ks) earlier observation 
of NGC 3783 by \xmm\ have been published by \citet{Blustin02}, who
highlight the complexity of the iron line profile in this object, whilst 
an analysis of the RGS spectrum from this observation have recently been 
presented by \cite{Behar03}. Together with the long 
300 ks \xmm\ exposure of the Seyfert 1 MCG -6-30-15 \citep{Fabian02}, 
the current observation represents 
the best available dataset with which to study the iron K line 
profile in AGN, as the long \xmm\ observations offer very high signal to 
noise up to 12 keV. This allows us to study the various components, such as 
the narrow and broad lines, the reflection hump above 
7 keV and any absorption lines or edges present in the iron K-shell band. 
The long \xmm\ observation also makes it feasible to probe any changes in the 
iron K profile and the continuum on relatively short timescales. 
The higher resolution, but lower signal to noise 
900 ks Chandra-HETGS spectrum complements 
this \xmm\ observation by allowing the study of  
the narrow emission/absorption lines
at higher spectral resolution; detailed modeling of the iron K band in 
this non-simultaneous dataset will be presented in another paper 
(Yaqoob \et 2003, in preparation).  

In Section 2, the \xmm\ observations of NGC 3783 are outlined, whilst in 
Section 3 the detailed modeling of the time-averaged iron K line profile 
is performed. In Section 4 the effect of the warm absorber 
on the iron line profile is investigated, whilst in Section 5
we discuss the variability of the iron K-shell 
band and continuum over the observation. 

\section{The XMM-Newton Observations}

NGC 3783 was observed by \xmm\ between 17-21 December 2001 over 2 complete 
satellite orbits (OBSID 0112210501 and 0112210201), 
with a total good exposure of just over 240 ks with the 
EPIC (European Photon Imaging Camera) detectors. 
Data was taken with the EPIC-pn detector 
\citep{Struder01} in Small Window Mode and with the EPIC-MOS detectors 
\citep{MTurner01} in Full Window mode and timing mode. The data was
reduced using version 5.4 of the XMM-SAS software using the standard 
processing scripts (\textsc{emchain} and \textsc{epchain}). 
Only short time intervals were excluded during the end of each satellite 
orbit, due to high count rate background flares, the background rates were 
nominal for the remainder of the observation.
Data were selected using event patterns 0-12 
(for the MOS) and pattern 0-4 (for the pn) and only good X-ray events 
(using the selection expression `FLAG=0' in \textsc{evselect}) 
were included. The spectra were extracted from circular
source regions of 40\arcs\ radius, whilst background spectra were
extracted from an offset circle of identical size, 
close to NGC 3783, but free of any
background sources. Response matrices and ancillary response
files were generated using the SAS tasks \textsc{rmfgen} and 
\textsc{arfgen} respectively.
The time-averaged, 0.2-12 keV flux of NGC 3783 during the observation was 
$6.8\times10^{-11}$~ergs~cm$^{-2}$~s$^{-1}$. The EPIC-pn lightcurve extracted 
over the full 0.2-12.0 keV band for both orbits is shown in Figure 1.

Unfortunately none of the 
MOS data taken during the observation are suitable for detailed spectral 
analysis, as the full window observations are heavily piled up at the flux 
level of the source, whilst the MOS timing modes are not presently 
calibrated to sufficient accuracy. Conversely the EPIC-pn exposures taken 
in Small Window mode do not suffer from significant pile-up, at the level of  
$<2$\% of events. Thus the spectral analysis was 
restricted to the EPIC-pn detector, which provides the highest 
signal to noise ratio in the iron K-shell band up to 12 keV.
Background subtracted spectra were fitted using \textsc{xspec} v11.2,
including data over the energy ranges 0.3 to 12 keV.
A Galactic absorption column of $N_{H}=8.5\times10^{20}$~cm$^{-2}$ 
\citep{Dickey90} was included in all the fits 
and fit parameters are quoted in the rest-frame of NGC 3783 at z=0.00973.
Given the large number of counts available in the observation, 
the source spectra were binned to a minimum of 50 counts per bin to enable 
the use of the $\chi^{2}$ minimization process when performing X-ray 
spectral fits.  
All errors are quoted at 90\% confidence for one 
interesting parameter (corresponding to $\Delta\chi^{2}=2.7$).

\subsection{Initial Spectral Fits}

Initially we concentrate our analysis on the time-averaged 
NGC 3783 spectrum, from the whole 240 ks observation, 
which provides us with the highest signal to 
noise, especially important for studying the iron K-shell band in the 
greatest detail. A time dependent analysis 
is described in Section~5, which shows that any temporal spectral changes
are subtle, and do not effect our conclusions about the 
mean iron line profile.
Initially a single power-law was fitted over a relatively clean 
part of the EPIC-pn spectrum 
between 3.5-5 keV, i.e. avoiding the iron line above 
5 keV and the strong warm absorber in the soft X-ray part of the spectrum. 
The best-fit photon index was $\Gamma=1.6$, whilst 
the spectrum and data/model residuals (Figure 2)  
show a clear deficit of counts between 0.7 and 3 keV, due to the 
known soft X-ray warm absorber in NGC 3783, whilst strong residuals are also 
present above 6 keV in the iron K-shell emission/absorption band. A weak 
soft X-ray excess is also present at the lowest energies below 0.7 keV, 
which has also been detected in an earlier \sax\ observation 
\citep{DeRosa02}.

In order to model the iron K shell band we concentrate 
our analysis on the higher energy portion of the spectrum. The exact value of 
the low energy cut-off used is important, as the aim is to analyze part of the 
spectrum which is largely unaffected by the strong soft X-ray warm absorber, 
which can have a considerable effect in the determination of the underlying 
continuum. Inspection of Figure 2 shows that the absorber 
starts to have a significant effect below 3 keV in the EPIC-pn data. 
We have also studied the (non-simultaneous) 900 ks Chandra-HETGS 
observation presented by \citet{Kaspi02}. All the strong absorption 
features in this spectrum are present below 4.5 Angstrom (2.8 keV) mainly 
due to K-shell absorption lines from elements between C and S, as well as 
L-shell absorption from iron. The last abundant element 
that may contribute 
discrete spectral features in the soft X-ray absorber is Ar, the 
HETGS spectrum shows that there are weak 
absorption lines due to He and H-like Ar at 3.1 and 3.3 keV respectively. 

In order to minimize the effect of the warm absorber when analyzing the 
iron K line profile, 
we adopt a conservative approach, and restrict our spectral 
analysis to the energy range from 3.5 to 12 keV. Over this band the only 
discrete absorption features (other than from iron) 
are due to Ca~\textsc{xix} and Ca~\textsc{xx}. These features are 
weak in the 900 ks HETGS observation, although note that the S/N of the 
HETGS above 3.5 keV is reduced, compared to the soft X-ray portion of the 
spectrum. In a later section of this paper (Section 4) 
we also discuss the effect the warm absorber 
may have on the continuum (through bound-free absorption) 
and above the neutral iron K-shell edge,
by comparing our fits with models generated with the 
photo-ionization code \textsc{xstar} \citep{Kallman96}. 

\section{The Time-Averaged Iron line Profile of NGC~3783}

Initially we fitted the hard band (3.5-12 keV) EPIC-pn spectrum with a 
simple power-law together with neutral absorption from our own 
Galaxy. The fit was very poor ($\chi^{2}/dof=3719/1292$), with a best-fit 
photon index of $\Gamma=1.58\pm0.01$. All the fit parameters are listed in 
Table 1. The data/model residuals are plotted 
in Figure 3 (panel a), two strong emission lines are apparent close 
to 6.4 keV and 7.0 keV, as well as a deficit of counts both near 
6.6-6.7 keV and above the neutral iron K edge at 7 keV, whilst a small 
excess of counts is observed red-wards of the 6.4 keV line. 

As a first step, we model the two strong emission lines with simple 
Gaussian profiles (Table 1, fit 1), 
with the energy, width and line flux free parameters 
(in addition to the continuum photon index and normalization). The line 
centroids are at $6.39\pm0.01$~keV and $7.00\pm0.02$~keV, with equivalent 
widths of $123\pm6$~eV and $34\pm5$~eV and fluxes of 
$(5.9\pm0.3)\times10^{-5}$~erg~cm$^{-2}$~s$^{-1}$ and 
$(1.4\pm0.3)\times10^{-5}$~erg~cm$^{-2}$~s$^{-1}$ respectively. The $1\sigma$ 
widths of the 6.4 and 7.0 keV lines are $57\pm8$~eV and $53\pm23$~eV 
respectively, which 
correspond to FWHM velocity widths of $\sim6100$~km~s$^{-1}$ and 
$\sim5200$~km~s$^{-1}$ respectively. Note that the non-zero widths 
of the lines 
are {\it required} by the data, for the $K\alpha$ line the 
improvement in fit statistic is $\Delta\chi^{2}=81$ for allowing the width 
to be non-zero 
(corresponding to an F-test null hypothesis probability of 
$1.3\times10^{-15}$). We comment on possible origins of 
the width of the 6.4 keV line later.

The observed ratio of the line fluxes are approximately 4:1, whereas one 
would expect a ratio of 150:17 between the $K\alpha$ and 
$K\beta$ transitions from neutral iron. Thus there is likely to be an 
extra component contributing towards the emission line at 7.0 keV.
Indeed the line energy of the 7.0 keV 
line is in between the lab frame energies 
for Fe~\textsc{xxvi} Ly$\alpha$ and Fe~\textsc{i} K$\beta$, which 
indicates that this line is a blend of these two emission lines
(and which may 
in part account for the velocity width of this line component). 
In order to determine the strength of the putative H-like iron line, 
we fixed the ratio of the intensity of the $K\beta$ line at 7.06 keV to 17/150 
of that of the Fe K$\alpha$ line at 6.4 keV.  We also tied the velocity 
widths of the all 3 emission lines to that of the 6.4 keV line 
(Table 1, fit 2). 
Indeed a H-like Fe line component is 
required in the fits ($\Delta\chi^{2}=52$ for 2 extra degrees of freedom), 
the measured energy of $6.96\pm0.02$~keV is very close to the 
known rest-frame energy for Fe~\textsc{xxvi} at 6.966~keV 
\citep{Pike96}, whilst the equivalent width of the line is $20\pm5$~eV. 

Overall the fit statistic for this three component emission line fit (with 
$\chi^2/dof=1591/1287$) is not formally acceptable, corresponding to a 
null hypothesis probability of $1.15\times10^{-8}$. The data/model residuals 
to this fit are shown in Figure 3b, clearly there is a broad excess (red-wing) 
below 6.4 keV, whilst there appears to be an absorption line in the data 
near 6.6-6.7 keV. A strong Compton reflection component may also be present, 
as indicated in the spectrum by an edge above 7 keV and 
spectral hardening up to 12 keV. 
Indeed one would expect a Compton hump to accompany the 
6.4 keV iron K emission line, if the line results from reprocessing in 
Compton-thick matter. Thus a neutral Compton reflection component, 
the \textsc{pexrav} model in \textsc{xspec} \citep{MZ95}, 
was added to the model. The 
strength of the reflection component (measured by the parameter 
$R=\Omega/2\pi$, where $\Omega$ is the solid angle in steradian covered 
by the reflecting material) was initially fixed at $R=1$, 
with an inclination of 30\degg and a cut-off energy 
of $\sim250$~keV using solar abundances. Note these parameters are consistent 
with the measurement of a Compton hump in an earlier \sax\ observation 
\citep{DeRosa02}, where the strength of the reflection component 
was measured to be $R\sim0.8$ with a cut-off energy of about 300~keV. 
After adding the Compton reflection component, the fit 
statistic improves considerably ($\chi^{2}/dof=1492/1287$), 
whilst all 3 emission 
lines are still required in the dataset (Table 1, fit 3). The underlying 
power-law slope is now steeper ($\Gamma=1.73\pm0.01$), accounting for 
the soft excess (e.g. Figure 2) 
observed in the broad-band EPIC-pn spectrum. After allowing  
the strength of the reflection component to vary, 
we obtain $R=1.6\pm0.3$, with 
a steeper photon index ($\Gamma=1.78\pm0.01$). However as the exact 
value of $R$ is very dependent on the high energy calibration of the 
EPIC-pn above 8 keV, we proceed by fixing R to 1, which is 
consistent with the strength of the Fe K$\alpha$ line measured in the 
spectrum \citep{George91} and the measurement of the 
reflection hump by \sax\ \citep{DeRosa02}.

After adding the reflection component to the spectral fit  
the excess of counts below 6.4 keV and the absorption line 
near 6.7 keV are still apparent in the residuals (Figure 3c). To 
model the apparent red-wing to the Fe K$\alpha$ line we added a `Diskline' 
component \citep{Fabian89}, to represent the emission from the 
inner accretion 
disk around a Schwarzschild black hole, taking the inner and outer 
radii of the disk as $6R_{S}$ and $100R_{S}$ respectively (where 
$R_{S}=2GM/c^{2}$) and fixing the rest-energy of the diskline emission 
at 6.4 keV. The improvement in fit statistic upon adding the diskline 
component is substantial ($\Delta\chi^{2}=112$ for 3 additional parameters). 
The emissivity ($\beta$) of the diskline 
(where the disk emission varies with radius as $R^{-\beta}$) 
was $\beta=3.3\pm0.5$, the inclination angle derived was $19\pm9$\degg\ whilst 
the strength of the line is relatively weak (the equivalent width is 
$58\pm12$~eV). 

The spectral fit was improved further by the addition of a 
narrow (unresolved by EPIC-pn) 
absorption line at $6.67\pm0.04$~keV, with an equivalent 
width of $17\pm5$~eV.  Note the other iron emission line components 
are required as in the previous fits (Table 1, fit 4). 
The absorption line component is unambiguously required in the model fit
with a high degree of statistical significance ($\Delta\chi^{2}=56$ for 
2 additional degrees of freedom), equivalent to an F-test 
null hypothesis probability of 
$2.9\times10^{-12}$. The absorption line cannot be modeled with another 
emission component, 
such as the blue-wing of the disk emission line for instance, as the 
line is observed below the level of the power-law continuum (e.g. see 
Figure 3, panel c and d).
The energy of this absorption line component 
probably corresponds to a blend of several lines of highly ionized iron,   
e.g. Fe~\textsc{xxiii} at 6.62 keV, Fe~\textsc{xxiv} at 6.66 keV and 
Fe~\textsc{xxv} at 6.70 keV. 
Overall the fit statistic ($\chi^2/dof=1324/1282$) is 
now formally acceptable, no other spectral components are required.
 
Finally we have also addressed the issue of whether the calibration of the 
EPIC-pn detector in the iron K-band has any effect on the fits, especially 
as some of the iron K features, discussed above, are present at the 5-10\% 
level above the continuum. For comparison, we have therefore reduced the 
spectrum of the bright BL-Lac object, Mrk 421, obtained during a calibration 
observation with the EPIC-pn in small window mode on 14 November 2000. 
The featureless continuum spectrum was fitted with a 
simple absorbed power-law above 3.5 keV with a best-fit power-law 
index of $\Gamma=2.65\pm0.01$. The data/model ratio residuals to this 
power-law model are shown in Figure 4, clearly no strong systematic features 
are observed over the iron K band, to about the 2\% level over the 
power-law continuum. Thus the EPIC-pn calibration is unlikely to effect the 
iron K profile modeling in NGC 3783, as the Fe K emission and absorption 
present appears stronger than any systematics due to calibration over this 
bandpass. 

\subsection{The Width of the 6.4 keV Fe line}

After modeling all the emission and absorption line components, and taking 
into account the resolution of the EPIC-pn detector (through the pn 
re-distribution matrix), the best fit $1\sigma$ width of the 
6.4 keV K$\alpha$ line is $\sigma=52\pm10$~eV. 
In terms of the FWHM width, this corresponds to $120\pm23$~eV or 
$5600\pm1100$~km~s$^{-1}$ and appears to be just resolved by the 
EPIC-pn detector (see Figure 5). Note that the neutral Fe K$\alpha$ 
line is actually a 
blend of two lines, at 6.391~keV and 6.404~keV, with a branching 
ratio of 2:1. However taking this into account in the spectral fits has a 
negligible effect on the velocity width of the line. 

We note that this 
velocity width is broader than that reported by \citet{Kaspi02} from the 
900~ks Chandra HETGS observation. However some caution should be exercised 
when interpreting the velocity width obtained from the lower resolution 
EPIC-pn spectrum (FWHM resolution $\Delta E\sim120$~eV at 6 keV) 
compared to the 
higher resolution HETGS spectrum ($\Delta E\sim35$~eV at 6 keV). 
For instance the 
6.4 keV line may appear to be slightly broader, as the EPIC-pn is largely 
unable to resolve the first Compton scattering shoulder of this line 
(at 6.24 keV), which is apparent in the 
HETGS data (Kaspi \et (2002), Yaqoob \et 2003, in preparation). 
In addition, some contribution 
from a weak outer disk emission line component may also contribute towards the 
width of the EPIC-pn line. Thus the quoted velocity width of 
$\sim5600$~km~s$^{-1}$ should be regarded as an upper limit. The effect 
of the Compton shoulder is further discussed in section 6.1.

\section{The Effect of the Warm Absorber}

Whilst measuring the iron K line profile is our primary aim in this 
paper, it is important to verify whether the deep soft 
X-ray warm absorber in NGC 3783 absorber has any effect on the high 
energy spectrum and the iron line profile. The warm absorber could introduce 
subtle spectral curvature below 6.4 keV, and there may be some 
additional opacity above the neutral iron K-shell edge at 7.1 keV. Whilst 
this will have little effect on modeling the narrow Fe emission line 
components, the effect on the weak, broad red-wing (which constitutes 
only 5\% of the continuum at 6 keV) may be crucial. Furthermore 
there is direct evidence for a high ionization component of the warm absorber 
in the iron K-shell band, through the detection of a strong 
6.7 keV absorption line.

As a starting point, we base our initial models on those of \citet{Blustin02}, 
who fit the RGS spectra from 
an earlier, short (40 ks) \xmm\ observation of NGC 3783. Conveniently, 
these authors parameterized their model in terms of a two zone warm absorber; 
a high ionization component responsible 
for the He and H-like K-shell absorption lines from the abundant soft X-ray 
elements (e.g. C, N, O, Ne, Mg) as well as L-shell absorption 
from highly ionized iron (Fe~\textsc{xvii-xxiv}),
with a low ionization component responsible for the unresolved 
transition array (UTA) resulting from a blend of inner-shell 
$2p \rightarrow 3d$ absorption lines from Fe M-shell ions. 
Although this model 
parameterizes an earlier \xmm\ observation, 
it is thought that the warm absorber 
in NGC 3783 is relatively stable between the two \xmm\ observations, 
as shown recently 
by \citet{Behar03}, who argue that the most recent RGS spectra
from the long \xmm\ look agree well with this earlier 
\citet{Blustin02} model. Additionally no variability was found 
in the warm absorber 
\citet{Netzer03} from a detailed analysis of all the Chandra observations 
over time of this source. Note however, that over a 3 year timescale, 
\citet{George98} did find a factor of 2 change in the column density of the 
absorber in ASCA data. 

In order to duplicate this soft X-ray absorption as closely as possible, 
we generated 
two grids of \textsc{xstar} photoionization models with similar parameters 
to those used in the \citet{Blustin02} model. The relative 
elemental abundances of C through to Fe 
quoted in \citet{Blustin02} were used, together with an outflow velocity 
of 800~km~s$^{-1}$ and a ($1\sigma$) turbulence velocity of 500~km~s$^{-1}$. 
As per \citet{Blustin02}, for the 
low ionization component we use an iron over-abundance of 10 times solar, 
and then fix the column density and ionization parameter at 
$N_{\rm H}=6\times10^{20}$~cm$^{-2}$ and 
log~$\xi=0.3$~erg~cm~s$^{-1}$ respectively.
Note however that this low ionization component is only added for 
completeness, it has relatively little effect on the 
high energy spectrum above 3.5 keV, only adding a small amount of 
opacity near the neutral iron K edge at 7.1 keV. 
For the higher ionization component, we initially fix the column density 
at $N_{\rm H}=3\times10^{22}$~cm$^{-2}$ with an ionization parameter of 
log~$\xi=2.4$, using solar abundances of iron.

This two component 
model was then applied to the EPIC-pn spectrum over the 3.5-12 keV 
bandpass. Two narrow emission line components (one representing 
Fe K$\alpha$, the 2nd representing the blend of Fe K$\beta$ and 
Fe~\textsc{xxvi}) were included in the fit, 
as well as the reflection component 
(with R fixed at 1) and the power-law continuum. 
Neither the diskline component 
nor the absorption line at 6.7 keV were included in the model. 
Initially the fit is 
poor ($\chi^{2}/dof=1528/1287$), the model does not reproduce the energy and 
depth of the strong 6.7 keV absorption line present in the data, as the 
ionization state of even the higher ionization absorber 
is not sufficient to produce the required column of highly ionized 
iron (e.g. Fe~\textsc{xxiii}--\textsc{xxv}). 

Thus, to fit the Fe K-shell absorption line present in 
the \xmm\ EPIC data,  
we allowed the ionization parameter and column density of the highest  
ionization absorption component to vary. A 
higher ionization parameter of 
${\rm log}~\xi=2.9\pm0.3$~erg~cm~s$^{-1}$ 
is required, whilst the column density 
is now $N_{\rm H}=(4.6\pm0.8)\times10^{22}$~cm$^{-2}$ and the 
underlying continuum slope is $\Gamma=1.69\pm0.01$. The fit to the spectrum 
has significantly improved ($\chi^{2}/dof=1332/1284$), reproducing the 
depth of the 6.7 keV absorption line well (see Figure 6), 
as the ionization state of the absorber is now high enough for 
Fe~\textsc{xxiii}--\textsc{xxv} to be the dominant ions present. 
Thus this probably represents the highest ionization component of the 
known warm absorber in NGC~3783. 

Interestingly, with the addition of the warm absorber, there 
is now no longer any requirement for the broad, relativistic disk emission 
component, the fit statistic is not significantly improved upon the 
addition of a diskline component, as the residuals to the warm absorber fit 
are small. (Note that the existence of the narrow Fe emission components 
are not affected). Formally the 90\% upper-limit on a disk emission line 
component is $<35$~eV, assuming a disk emissivity of $\beta=3$ and 
an inclination angle of 30\degg.  However note that the iron abundance 
of the high ionization absorber may be super-solar, hence the required 
hydrogen column density could be lower. For an iron 
abundance $10\times$ solar, the required hydrogen column will then be 
$4.6\times10^{21}$~cm$^{-2}$. In this scenario, then amount of spectral 
curvature below 6 keV due to the warm absorber is lessened and 
it is not possible to totally exclude the presence of 
a broad iron K line. Nonetheless any such broad line is very weak, 
with an equivalent width of only $23\pm12$~eV, whilst the improvement in the 
fit statistic upon adding the line is marginal 
($\Delta \chi^2=12$ for 2 extra parameters).

In order to assess directly the effect of the high ionization absorber on the 
spectrum above 3.5 keV, we removed the warm absorber from the 
model. The result is shown in Figure 7, clearly one can see the contribution 
that the high ionization absorber makes towards the 6.7 keV absorption line. 
However the absorber also introduces some continuum curvature
below the iron K line (which effects the fits to the broad line component). 
There is also some opacity between 7-9 keV due to iron K-shell bound-free 
absorption, with an optical depth $\tau \lesssim 0.1$. 
This is also illustrated 
in Figure 8, which plots the highest ionization component of the 
warm absorber model, normalized 
to a $\Gamma=2$ power-law continuum. One can clearly see the continuum 
curvature between 3-6 keV, due to recovery from highly ionized K-shell 
edges from Mg, Si and S as well as from the 
L-shell edges of Fe below 3 keV. Additional opacity 
is also present above 8 keV, 
due to a blend of iron K-shell edges, as well the K$\beta$ 
(and higher series) absorption lines of highly ionized iron.

Finally we extrapolated this warm absorber to lower energies. 
However, to model the broad 
band spectrum (from 0.3-12 keV), we include 
three different ionization components, the two lower ionization 
(with log~$\xi = 2.4$ and log~$\xi = 0.3$)
warm absorber components used by \citet{Blustin02}, which fit the earlier RGS 
spectrum, as well as the highest ionization component (with 
log~$\xi = 2.9$) ressponsible for the 6.7 keV iron absorption line. 
Allowing the ionization parameters and columns of the 3 component 
warm absorber to vary then gave log~$\xi$ of -0.1, 2.1 and 
3.0 erg~cm~s$^{-1}$, with column densities of 
$1.1\times10^{21}$~cm$^{-2}$ (for $10\times$ solar Fe abundnace), 
$1.2\times10^{22}$~cm$^{-2}$ (solar Fe) and 
$4.4\times10^{22}$~cm$^{-2}$ (solar Fe) for the lowest to highest 
ionization components of the absorber respectively. The overall 
fit statistic was $\chi^{2}/dof=2623/1893$, whilst the broad-band spectrum, 
and data/model residuals to the warm absorber model, are plotted in 
Figure 9. The remaining residuals present 
in the spectrum are small, at the 5\% level, similar to the 
residual calibration uncertainties in the EPIC-pn responses in the soft X-ray 
portion of the spectrum. 
The continuum 
photon index returned was $\Gamma=1.67\pm0.01$, whilst a weak soft excess 
modeled by a black-body with a temperature of $kT=97\pm4$~eV 
was required to fit the very softest part of the spectrum. 

Note that all 3 
warm absorber components are required to model the full band 
X-ray spectrum. For instance, removal of the highest ionization absorber 
results in a significantly  
worse fit, with $\chi^2/dof=2696/1895$, as the 6.7 keV absorption line 
is then not modeled by the absorber. Thus whilst the two lower ionization 
components provide a satisfactory fit to the spectrum below 2 keV (i.e. 
as shown by \citet{Blustin02}), a third higher ionization component is 
required to model the 6.7~keV iron K-shell absorption. Indeed \cite{Netzer03} 
require at least three different warm absorber components to model the 
900 ks Chandra spectrum. It is therefore 
plausible that a wide range of ionizations are present in the absorber in 
NGC 3783; for instance iron with ionization states $<$~\textsc{xvii} is 
required from the detection of the iron M-shell UTA in the RGS spectrum 
\citep{Blustin02, Behar03}, whilst Fe~\textsc{xxv} is likely to 
contribute towards the absorption at 6.7 keV.

\section{Variability of the Fe line and hard X-ray continuum}

Initially we split the observation according to the two separate \xmm\ orbits 
(e.g. Figure 1). The time-averaged 
flux state was relatively low during the first orbit 
(3.5-10 keV band flux of $2.77\times10^{-11}$~erg~cm$^{-2}$~s$^{-1}$) and 
higher during the 2nd orbit 
(3.5-10 keV band flux of $3.69\times10^{-11}$~erg~cm$^{-2}$~s$^{-1}$). 
As in the earlier fits, the iron K-shell spectral features were parameterized 
by multiple Gaussian emission/absorption line components and a neutral 
reflector (with $R=1$). A slight steepening with flux of the hard 
power-law index was observed, 
from $\Gamma=1.62\pm0.01$ in the first orbit to $\Gamma=1.68\pm0.01$ 
during the 2nd orbit. However there was no change in the 6.4 keV Fe $K\alpha$ 
line parameters, the Fe line flux being consistent with a constant 
value ($5.7\pm0.4\times10^{-5}$~photons~cm$^{-2}$~s$^{-1}$ during the 
first orbit and $5.4\pm0.5\times10^{-5}$~photons~cm$^{-2}$~s$^{-1}$ during 
the 2nd orbit). These values are also consistent with the Fe K$\alpha$ flux 
measured in the long Chandra HETGS observation, of 
$5.3\pm0.6\times10^{-5}$~photons~cm$^{-2}$~s$^{-1}$ \citep{Kaspi02}.
Note, given the much poorer statistics on the 7.0 keV line, 
it is not possible to determine whether or not this line component varied. 

However, a significant change in the 6.7 keV absorption line was observed 
from one orbit to the next, indicating that the highly ionized 
absorber is variable on timescales of $\sim10^{5}$~s. This is illustrated 
in Figure~10, which shows the ratio to the best fit model to the data 
(but not including the absorption component) for the two orbits. 
Thus the absorption line appears to be strongest during the second, 
higher flux orbit (with an equivalent 
width of $18\pm4$~eV), whilst it was barely detected during the first orbit 
(equivalent width of $7\pm4$~eV). 

The observations were also split into shorter time segments, of 
$8\times30$~ks duration (i.e. 4 segments per orbit). However the more limited 
photon statistics do not enable us to determine whether 
the 6.7 keV absorption line is variable over this shorter timescale. 
However a similar 
pattern was seen with regards to variability of the Fe K$\alpha$ line and 
the continuum. Whilst the 3.5-10 keV continuum flux varied by as much 
as 50\% during the observation, there was no change in the iron K$\alpha$ line 
flux (to $\pm$10\% of the mean value). Similarily, a change in continuum 
photon index was observed, steepening from $\Gamma=1.57\pm0.02$ at the lowest 
flux level to $\Gamma=1.71\pm0.02$ at the highest flux level. This is 
consistent with the well known positive 
Flux-Gamma correlation found in many active galaxies. An alternative 
interpretation is that 
the primary power-law index may infact be the same for the different 
flux states. However if the Compton reflector originates from distant 
matter and does not respond to the continuum (which appears 
consistent with the constant-flux K$\alpha$ emission line), then this 
component will appear stronger relative 
to the weaker power-law in the low flux spectra, resulting in a 
harder spectrum overall. 

\section{Discussion and Conclusions}

The long \xmm\ observation of the bright, Seyfert 1 galaxy NGC 3783 has 
revealed a complex iron K line profile above 3.5 keV. Two strong, but 
relatively narrow emission lines are apparent at 6.4 and 7.0 keV, the 
former from the (near) 
neutral Fe K$\alpha$ fluorescence line, whilst the latter 
is likely to be a blend of the neutral Fe K$\beta$ line as well as a 
component from hydrogenic iron (i.e. Fe~$\textsc{xxvi}$ Ly-$\alpha$). 
The observations also revealed a weak red-wing to the 
iron K$\alpha$ line profile below 6.4 keV, as well as an unambiguous 
detection of a high ionization iron absorption line at 6.7 keV. Note that 
at the resolution of the EPIC-pn, we cannot preclude the presence 
of emission or absorption from intermediate states of iron between 
6.4 and 6.7 keV.

\subsection{The Nature of the Iron K Line Emission in NGC 3783}

The 6.4 keV Fe K$\alpha$ line appears to be resolved in the 
EPIC-pn spectrum, with a typical (FWHM) velocity width of 
$\sim5000$~km~s$^{-1}$. At first glance this velocity is consistent 
with the FWHM dispersion in the BLR of NGC 3783 
\citep{Riechert94, Wandel99}. 
However we note that the velocity width determined by the higher 
spectral resolution Chandra HETGS observation is much lower, the core of the 
6.4 keV line 
having a FWHM width of $\sim1800$~km~s$^{-1}$ \citep{Kaspi02}. 

One possible explanation for this apparent 
discrepancy is the lower spectral resolution of the EPIC-pn detector compared 
to the HETG. For instance at the EPIC-pn resolution 
($\Delta E \sim120$~eV at 6 keV), it is difficult to resolve the narrow 
core of the line from a broader component, 
such as the Compton scattering shoulder of the line, which is 
apparent in the Chandra line profile (Kaspi \et (2002),  
Yaqoob \et (2003), in preparation). 
However one can substitute in the \xmm\ spectrum 
for the parameters found in the Chandra HETGS 
fits to the Compton shoulder. For instance Yaqoob \et 2003 approximate the 
Compton shoulder with a Gaussian profile, with a line energy of 6.24 keV 
(corresponding to the first Compton scattering peak, e.g. \citet{George91}), 
a line width of $\sigma=110$~eV and a line flux $\sim25$\% of the 
neutral K$\alpha$ line flux. If one takes the Compton shoulder of the line 
into account, then the broadening of the iron K$\alpha$ line in the \xmm\ 
spectrum is no longer 
required. The formal upper on the K$\alpha$ line width is then 
$\sigma<38$~eV, corresponding to a FWHM velocity width of $<4000$~km~s$^{-1}$.

Indeed the detection of the Compton scattering shoulder in the 
Chandra line profile, as well as the narrow width of the line core 
is indicative of scattering in distant Compton thick matter. 
Additionally both the \xmm\ and the earlier \sax\ (De Rosa \et 2002) 
observations also require a strong ($R\sim1$) Compton reflection 
component above 7~keV. However the line is unlikely to originate 
from the disk unless the typical radius is $>1000R_{s}$. Furthermore 
line does not respond to the continuum 
within the timescale (days) of the \xmm\ observation in December 2001, 
whilst the line flux 
also does not appear to vary within the 18 month timescale of the Chandra 
observations between 21 January 2000 and 26 June 2001 
(Yaqoob \et 2003, in preparation), or indeed between the 
Chandra observations and the December 2001 \xmm\ observation 
(the mean line flux from 
Chandra, of $5.3\pm0.6\times10^{-5}$~photons~cm$^{-2}$~s$^{-1}$, 
Kaspi \et (2002), is consistent with the \xmm\ value). Given the lack of 
variation on these timescales, 
the bulk of the 6.4 keV line would appear to originate from distant matter. 
In the context of AGN unification schemes \citep{Ant93}, one 
possible source for this distant, Compton-thick reprocessor is the 
putative molecular torus. Indeed predictions show that the putative 
torus may be a major contributer towards both the 6.4 keV line commonly 
seen in the Seyfert 1 spectra  from both \xmm\ and \chandra\ observations, 
as well as the strong iron lines observed in Seyfert 2s 
\citep{Ghis94}. Other geometries are also possible 
such as scattering off the Compton thick component 
of any quasar outflow \citep{Elvis2000}. 

Whilst the detection of a narrow and 
distant 6.4 keV iron line now appears robust in 
this and many of the Seyfert 1s, generally the presence of the 
broad iron K line 
from the inner accretion disk \citep{Tanaka95, Nandra97} is 
subject to considerable debate, the broad red-wing being much weaker than 
anticipated in the new \chandra\ and \xmm\ datasets, for instance in 
NGC 5548 \citep{Pounds03a} or NGC 4151 \citep{Schurch03}. 
If a disk-line component is present in NGC 3783 it is very weak, with an 
equivalent width of only 60~eV even before the warm absorber was modeled. 

However, once the high ionization absorber responsible for the Fe K-shell 
absorption is accounted for in NGC 3783, then the requirement for  
a broad line is further reduced, with a formal upper-limit of 
$<35$~eV on the equivalent width of such a component.
The reduction in strength of the broad line is mainly 
due to the fact that the absorber can introduce 
subtle continuum curvature in the X-ray spectrum, 
even in the iron K-shell band (e.g. see Figures 7 and 8). 
Indeed this current study may have implications 
for the detection of the broad iron K$\alpha$ line in other AGN, which also 
have a strong ionized absorbers (for instance in NGC 4151 or MCG -6-30-15). 
In the case of MCG -6-30-15, the broad line component is much stronger 
\citep{Tanaka95, Wilms01, Fabian02} and its detection appears to be 
more robust to the 
spectral model and underlying continuum slope that is assumed 
\citep{Reynolds03, Vaughan03}. However, the wealth of 
data now available through the \xmm\ and \chandra\ archives clearly 
call for a more 
thorough, systematic analysis of the iron line profile in many AGN.
 
Indeed an important question is why the broad line is not 
required in many of the 
\xmm\ Seyfert 1 spectra, and particularly in NGC~3783, which  
represents one of the highest quality 
iron line profiles obtained on any AGN to date? One possibility is that the 
inner disk is truncated, however given the narrow width of the 6.4 keV 
line, this requires that the bulk of the line 
emission occurs out at radii $>1000R_{s}$. However, at such a distance 
from the black hole, there is unlikely to be substantial hard X-ray emission.
Another scenario which is perhaps more realistic is that the inner disk is 
strongly photoionized, so that most of the iron at the disk surface 
is fully ionized, and the subsequently the line is 
too weak to be detected. The magnetic flare model where the X-ray flux 
is concentrated in small, intense regions above the disk, 
can produce a very 
highly ionized skin at the local disk surface, with a high Compton 
temperature, e.g. \citet{Nayakshin00}. 
The signature of this is a very weak iron line and disk reflection 
component, particularly when the underlying photon index is hard 
($\Gamma<2$). This scenario would appear consistent with the tight  
constraint on any broad disk emission line in these data, with an 
upper-limit of $<35$~eV on the line equivalent width. 

\subsection{The Origin of the Variable, Highly Ionized Iron Absorber}

Perhaps the most intriguing finding from this observation is the discovery of 
a variable absorption line component from highly ionized iron. 
The observed energy of the line 
($6.67\pm0.04$~keV) is consistent with the $1s \rightarrow 2p$ 
transitions of 
Fe~{\sc xxiii} (6.630~keV), 
Fe~{\sc xxiv} (6.659~keV) and 
Fe~{\sc xxv} (6.702~keV) 
at the systemic velocity of NGC~3783.
These transitions all have similar oscillator strengths 
($f_{osc} \simeq 0.6$--$0.7$), 
hence the observed feature could be a blend of these ions.
This absorber may represent the highest ionization phase of the 
gas that is responsible for the soft X-ray absorber in NGC 3783. 

However, recent studies indicate that the lower ionization absorption 
components in NGC 3783 \citep{Behar03, Netzer03} responsible for the
soft X-ray absorber, do not vary, implying that this absorbing matter 
is located at large, parsec scale distances. The detection of rapid 
variability in the iron absorber within the \xmm\ observation 
does not appear consistent with this; it is possible that the 
high ionization absorber is a physically separate component, 
located closer to the nucleus.
In order to provide a zeroth order estimate for the location of the 
high ionization absorber, we calculated 
the {\it maximum} possible distance to the iron absorber, on the 
condition that $\Delta R/R<1$, i.e. its thickness ($\Delta R$) cannot 
exceed its distance ($R$) from the nucleus. Combining the equations 
$N_{\rm H}= n \Delta R$ and $\xi=L/nR^{2}$ yields 
$R<L/N_{\rm H}\xi$. Now as $L=10^{43}$~erg~s$^{-1}$, 
$N_{\rm H}=5\times10^{22}$~cm$^{-2}$ and $\xi=10^{3}$~erg~cm~s$^{-1}$, then 
the maximum distance of the absorber is $2\times10^{17}$~cm, or $<0.1$~pc 
from the nucleus.

The \xmm\ observations show that the Fe 
absorption line is variable on timescales of 
$t_{\rm var} \lesssim 10^{5}$~s.
One possibility is that the variation in depth of the absorber 
between the two orbits may arise through changes in the 
ionization state of iron due to an increase in the illuminating flux.
In the optically-thin limit, 
Fe~{\sc xxiii}--{\sc xxv} are the dominant ions over a
wide range of $\xi$
($400 \lesssim \xi \lesssim 1600\ {\rm erg\ cm\ s^{-1}}$).
However the lack of significant absorption observed from the 
$1s \rightarrow 2p$ transition of Fe~{\sc xxvi} (6.966~keV)
implies an ionization parameter may be towards the lower end of this 
range.
For instance, if $\xi \sim 6\times10^{2}\ {\rm erg\ cm\ s^{-1}}$, 
then the relative ionization fractions
($f_{ion}$) will be 0.06, 0.40, 0.31 and 0.14 for 
Fe~{\sc xxvi}, Fe~{\sc xxv}, Fe~{\sc xxiv}, and Fe~{\sc xxiii}
respectively (and $f_{ion} << 0.05$ for the less-ionized species). 
With such an ionization structure, a reasonable fit to the absorption 
feature detected in the 2nd-orbit is obtained (see Figure 11, panel a)
for a total column density of 
Fe $N_{\rm Fe} \simeq 3\times10^{17}\ {\rm cm^{-2}}$
(corresponding to $N_{\rm H} \sim 10^{22}\ {\rm cm^{-2}}$
for an Fe abundance $A_{\rm Fe} = 2.7\times10^{-5}$, \citet{Wilms00}),
assuming ionization equilibrium has been reached.

Now let us consider the 1st-orbit, 
when the observed intensity of the source (in the 3.5-10 keV band) 
was a factor $\sim 1.3$ lower.
In the simplest case, when the absorbing gas is far from the nucleus
and sees the same continuum as ourselves, then the ratio of the 
illuminating flux (and hence $\xi$) between the 2nd- and 1st-orbits is 
also 1.3. However such a situation is rejected by the data: 
Fe~{\sc xxv}, Fe~{\sc xxiv}, and Fe~{\sc xxiii} ions would still be 
the dominant Fe ions  
(with $f_{ion} = 0.22$, 0.28 and 0.22, respectively) 
resulting in a predicted absorption blend stronger than observed.
However the 
observed hard X-ray continuum is likely to be a composite of 
the variable {\it primary} continuum from the central engine 
and a constant {\it reflected} continuum 
from distant matter. This reflection component is 
observed in the high energy \xmm\ 
spectrum, through the detection of a strong narrow K$\alpha$ line 
and a Compton hump above 7 keV; hence the flux 
change in the primary continuum emission may be higher, by as much as 
a factor of two, although the exact value will depend on the 
shape of the high energy continuum above 12 keV. 
Indeed a satisfactory model (Figure 11, panel b) 
can be obtained if we decrease the 
ionization state further in the first orbit 
by reducing ionization parameter to
$\xi \sim 3\times10^{2}\ {\rm erg\ cm\ s^{-1}}$, i.e. by a factor 
2 compared to that during the 2nd-orbit. Under these circumstances 
Fe~{\sc xxiii} and Fe~{\sc xxii} are the dominant ions (with 
$\lesssim 20$\% of the Fe in the form of Fe~{\sc xxiv} and above 
see Figure 11, panel c). 
Due to the lower oscillator strengths of Fe~{\sc xxii} and below, 
a weaker line is predicted which is in agreement with the data.

The above example calculations assume ionization 
equilibrium. One can then consider the circumstances under which 
this might be true. Specifically, one can calculate the 
ionization timescale ($t_{ion}$) required so that sufficient 
ionizing photons arrive and are absorbed by the gas such 
as to raise the ionization structure of the gas by the required 
amount. The column density of a given Fe ion is 
simply $N_{ion} = f_{ion} N_{\rm Fe}$ and for the 
underlying continuum 
appropriate for NGC~3783, 90\% of the ionizing photons 
will be absorbed within a band of width 
$\Delta E \simeq 0.8 E_{\rm th}$, where $E_{\rm th}$ is the 
threshold energy. For Fe~\textsc{xxiv}, 
the relevant threshold energy will be at 
2.05~keV, i.e corresponding to the L-shell edge 
energy of Fe~\textsc{xxiv}. 
If the number of photons emitted by the source at $E_{\rm th}$
is $N_{i}(E_{\rm th})$, and the distance to the absorber is $R$, 
then the number of photons absorbed (per unit area)
by a given ion within the gas 
in a time $t_{\rm ion}$ is 
$N_{i}(E_{\rm th}) \Delta E\ t_{\rm ion}\ (1 - e^{-\tau})/ (4 \pi R^{2})$,
where $\tau = \sigma_{\rm th} N_{ion}$ and $\sigma_{\rm th}$ is the 
cross-section at the threshold energy. 

Equating this to $N_{ion}$ then defines $t_{ion}$ for that ion.
Requiring $t_{ion} \leq t_{\rm var}$ then leads to an upper limit
on $R$ such that the ionization state of the absorber is able to 
react to variations in illumination within $t_{\rm var}$.
For the ions of interest noted above 
(i.e. Fe~{\sc xxi}--{\sc xxiv}), and assuming a distance of 42~Mpc 
towards NGC~3783, then the 
ionization equilibrium can be achieved in the $10^{5}$~s timescale 
between the 1st- and 2nd-orbits if 
$R \lesssim 0.02$~pc.
This is significantly smaller than the parsec-scale distances 
derived for the soft X-ray absorber \citep{Behar03, Netzer03}, but is 
consistent with the $\Delta R/R$ estimate above.
Note that for a warm absorber distance of $\sim0.1$~pc, the observed 
equilibrium time of $10^{5}$~s is consistent with the timescale 
calculated by \citet{KK01}, for Fe~\textsc{xxv}. 

Another possibility is that this absorber is similar to 
the extreme, high velocity ($v\sim0.1c$) 
iron absorbers recently observed in some AGN by \xmm\ and \chandra, 
e.g. APM ~08279+5255 \citep{Chartas02}; PG~1211+143 \citep{Pounds03c}; 
PG~1115+080 \citep{Chartas03}; PDS 456 \citep{R03}. 
These outflows may arise from the innermost part of a 
disk driven wind \citep{Proga00, King03}. In this case the variation 
seen in the Fe absorber may be due to the passage of ionized matter across 
the central X-ray source. 
For a given size of the X-ray emitting region 
in NGC~3783 one can estimate the distance of the transient absorbing 
material passing in front of the central engine, assuming that this occurs 
on the timescale of the variation seen in the iron absorption line, 
i.e. $\sim10^{5}$~s. In NGC~3783 the estimated black hole mass, 
obtained through BLR reverberation mapping, 
is $\sim10^{7}$~M$_{\odot}$ \citep{Onken02}, thus a typical size for the 
X-ray emitting region is $\sim20GM/c^{2}=3\times10^{13}$cm. 
Thus the velocity of the 
absorbing matter passing in front of the source in $10^{5}$~s 
is $v\sim3000$~km~s$^{-1}$ (or $0.01c$). If one equates this velocity 
to the escape velocity of the matter at a given radius R from the 
black hole, then the distance of the absorber is
$R\sim10^{16}$~cm. From the 
definition of the ionization parameter, 
$L_{\rm x}/\xi = nR^{2}$, whilst from observation 
$L_{\rm x}=10^{43}$~erg~s$^{-1}$ 
and $\xi=10^{3}$~erg~cm~s$^{-1}$, hence the density of 
the absorbing matter is $n\sim10^{8}$~cm$^{-3}$

The detection of the 6.7 keV absorption line in NGC 3783 
adds to the growing number of cases where high ionization iron 
K-shell absorption features have been detected.   
For those AGN where high velocity X-ray iron absorption features 
have been measured, a large mass outflow rate 
is often required (typically $>1$~M$_{\odot}$~year$^{-1}$), of the order of 
(or even greater than) 
the actual accretion rate required to power the bolometric luminosity 
of the sources, e.g. see the discussion in \cite{King03}. 
Similarly, we can also calculate the mass outflow rate required to power 
the high ionization absorber 
in NGC 3783, assuming an outflow velocity of $10^{3}$~km~s$^{-1}$ (i.e. 
consistent with the soft X-ray absorber). For a constant velocity outflow, 
the mass outflow rate will be:- 

$\dot{M}_{\rm out} = \Omega n R^{2} v m_{\rm p}$

where $\Omega$ is the solid angle subtended by the absorber and 
$m_{\rm p}$ is the proton mass. For the iron absorber parameters derived 
in NGC 3783 (and assuming $\Omega\sim\pi$~steradian), the mass outflow 
rate is $\dot{M}_{\rm out}=0.1 {\rm M}_{\odot}$~year$^{-1}$. This is 
of the same order as the expected accretion 
rate required to power the observed bolometric luminosity of NGC~3783
(assuming $L_{bol}=5\times10^{44}$~erg~s$^{-1}$ at 5\% accretion 
efficiency). Similar mass outflow rates in other Seyfert 1s, 
of up to one solar mass per year, were also calculated from ASCA 
observations, by \citet{Reynolds97}.
Note that if the soft X-ray component of the warm absorber does 
reside at parsec scale distances, then the mass outflow rate required 
to sustain this material is even higher, e.g. see the calculation in 
\citet{Behar03}, of the order 10~M$_{\odot}$~year$^{-1}$.  
If the outflows are a relatively persistent phenomenon, which is 
likely as the warm absorbers appear to reside in $>$50\% of Seyfert 
galaxies \citep{Reynolds97}, then the large mass outflow rates  
require that a significant proportion 
of the matter feeding the AGN may in fact be required to sustain the 
outflowing material.

\section{Acknowledgments}

This paper is based on observations obtained with XMM-Newton, an ESA
science mission with instruments and contributions directly funded by
ESA Member States and the USA (NASA). We would like to thank Richard 
Mushotzky and Simon Vaughan for discussions and feedback. We would also 
like to thank the anonymous referee for helpful comments. 
T.J. Turner and T. Yaqoob 
acknowledge support in the form of NASA grants, NAG5-7538 and NAG5-10769 
respectively.

\clearpage

\begin{figure*}
\rotatebox{-90}{
\epsscale{0.7}
\plotone{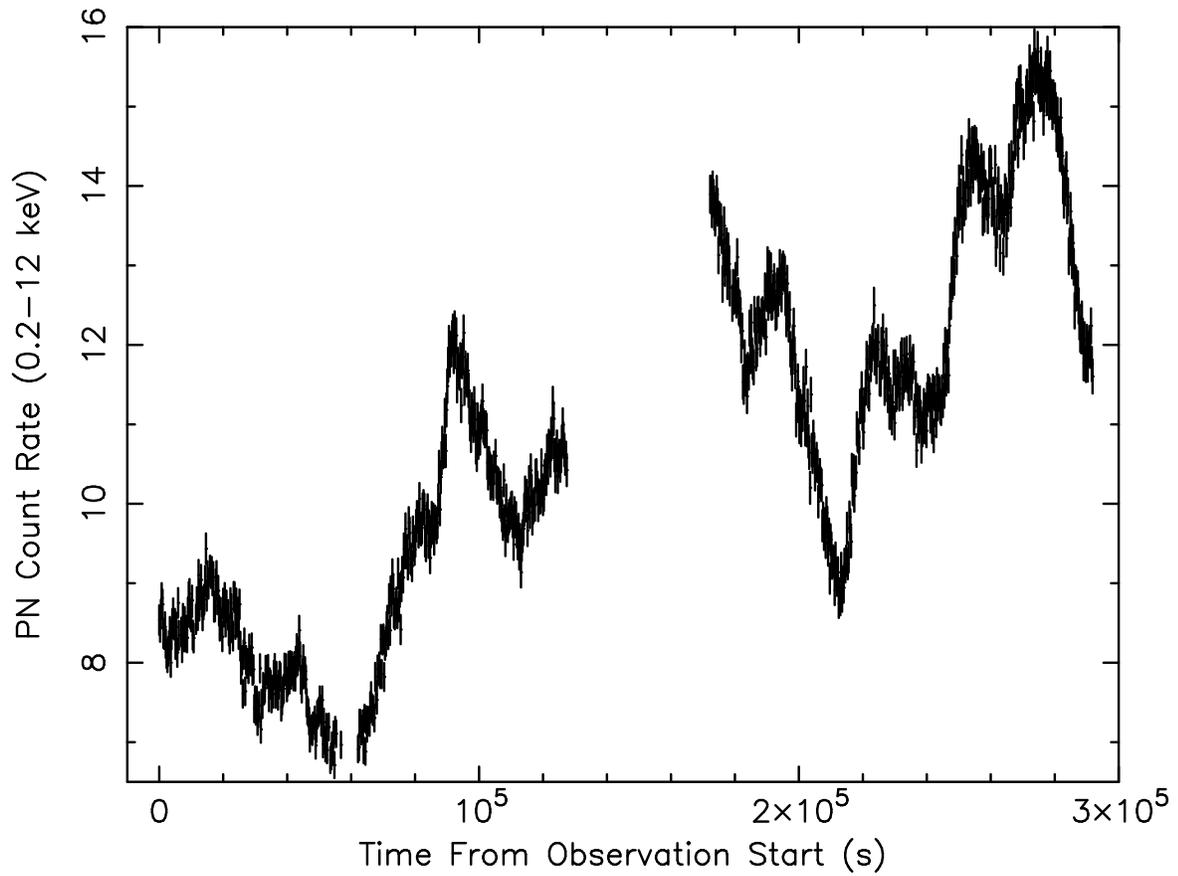}}
\caption{The 0.2-12 keV EPIC-pn lightcurve of NGC 3783, observed over two 
whole \xmm\ orbits. The observation started on 17 November 2001, with a total 
exposure time (after screening) of $\sim240$~ks.} 
\end{figure*}

\begin{figure*}
\rotatebox{-90}{
\epsscale{0.7}
\plotone{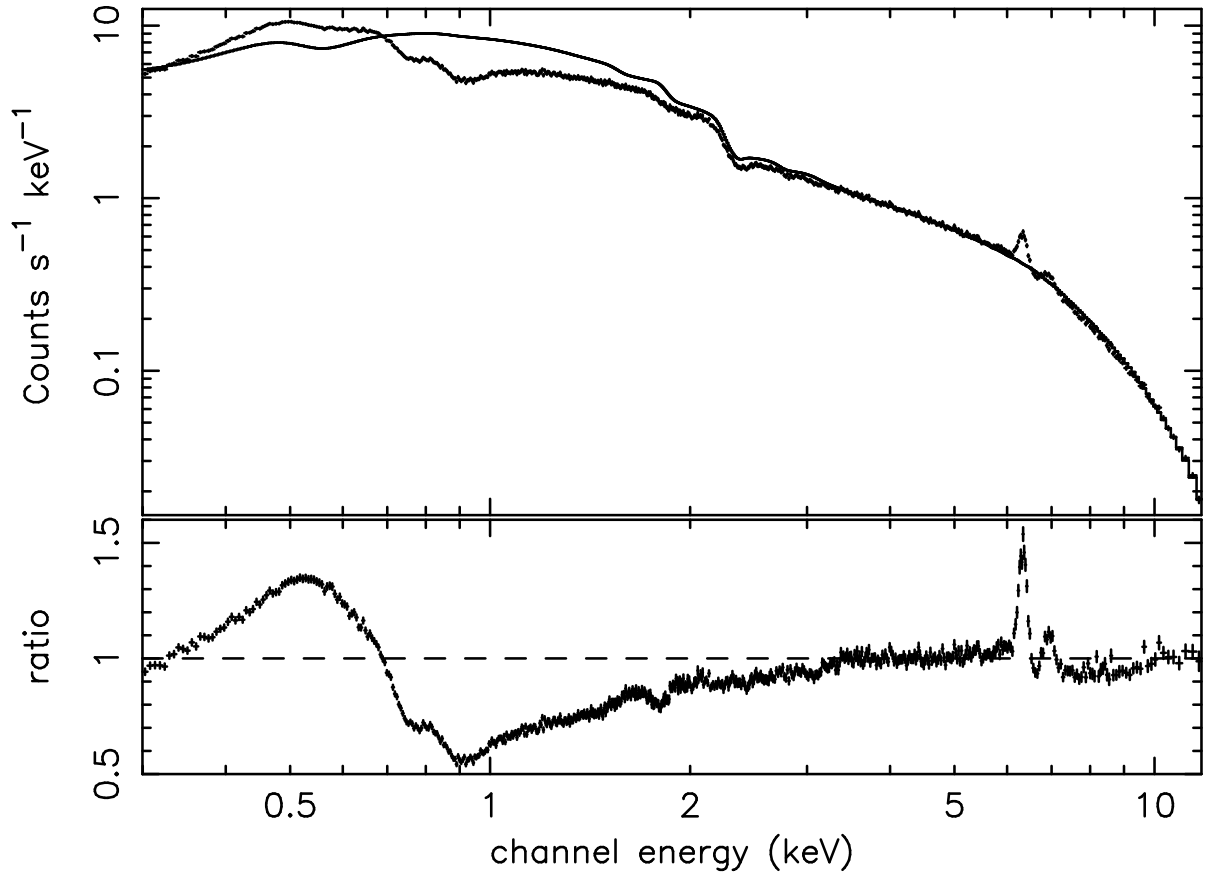}}
\caption{The broad-band (0.2-12 keV) EPIC-pn spectrum of NGC 3783. 
The upper panel shows the 
data, plotted against a power-law model of photon 
index $\Gamma=1.6$ (solid line), which has been convolved through the 
detector response matrix. The lower panel shows the data/model ratio 
residuals to this power-law fit. Clear deviations in the iron K-shell band are 
apparent between 6-7 keV, whilst a deficit of counts (due to the 
the warm absorber) is 
present between 0.7-3.0 keV, whilst a weak soft excess over the power-law 
continuum is observed below 0.5 keV.}
\end{figure*}

\begin{figure*}
\begin{center}
\rotatebox{-90}{\includegraphics[height=10cm]{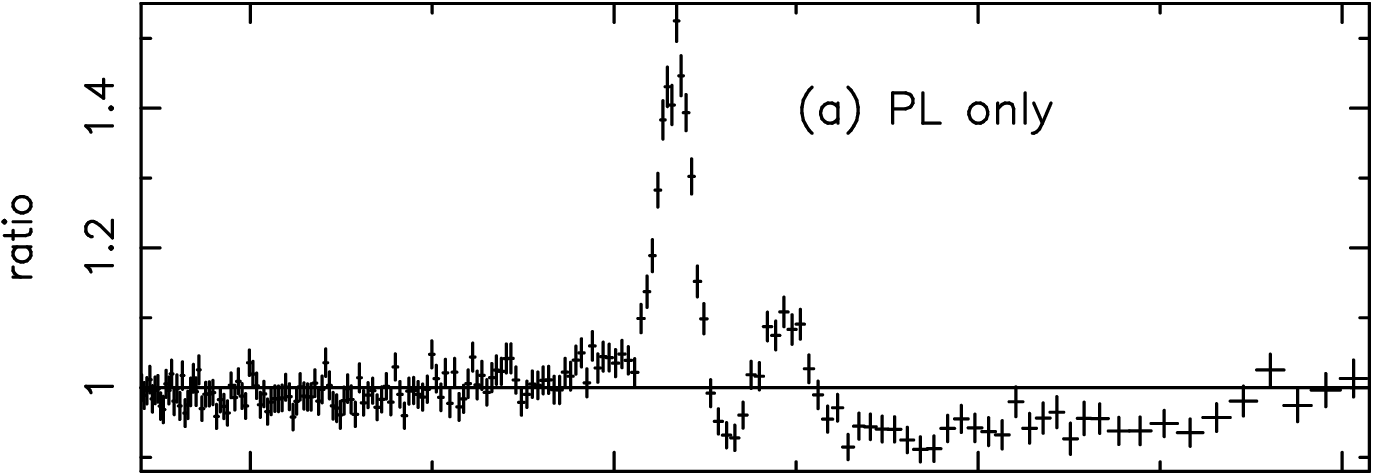}}
\rotatebox{-90}{\includegraphics[height=10cm]{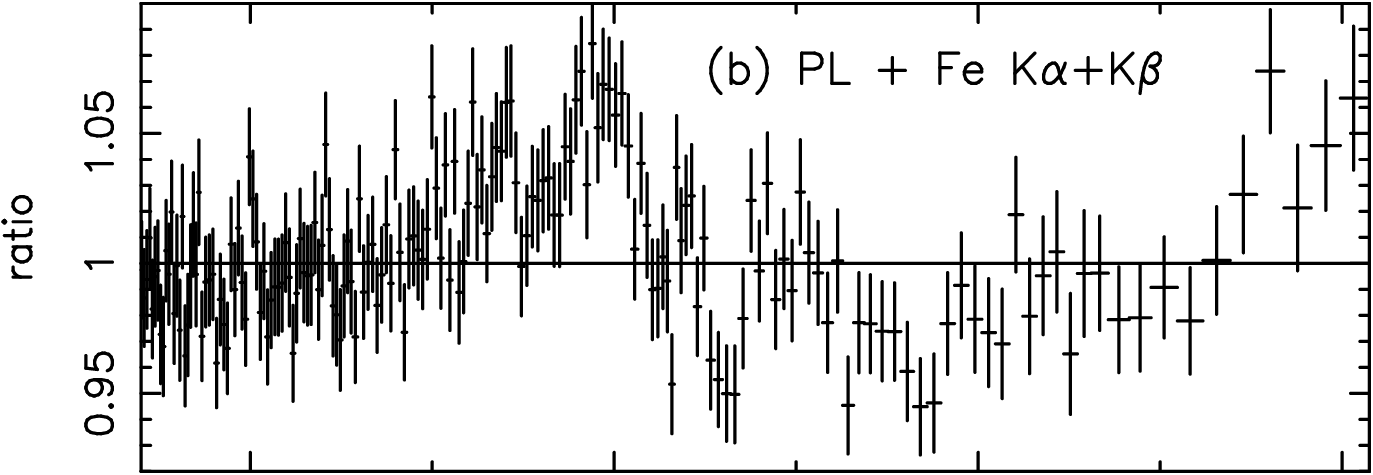}}
\rotatebox{-90}{\includegraphics[height=10cm]{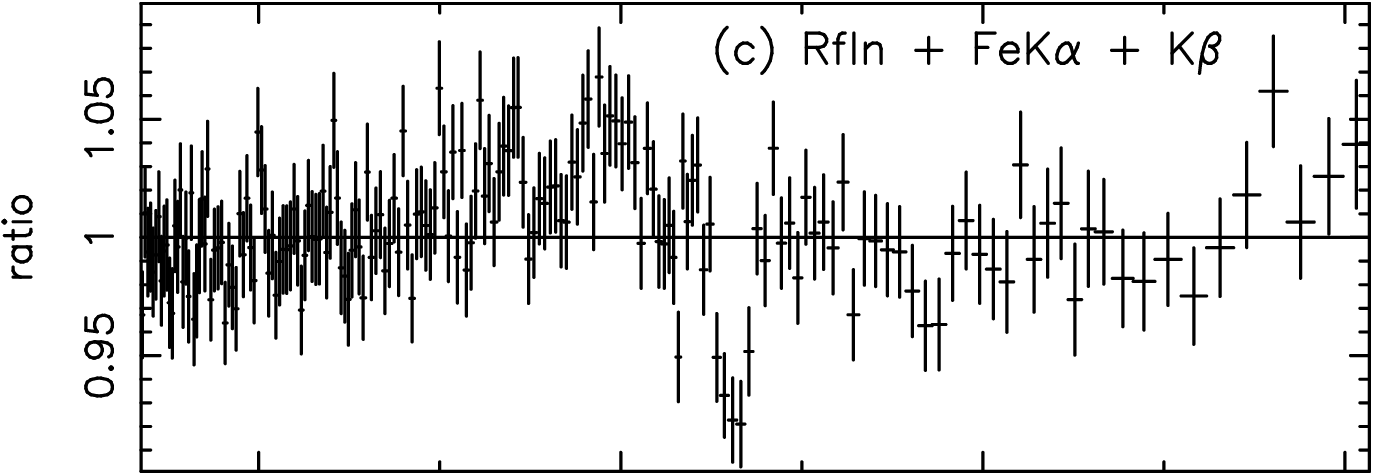}}
\rotatebox{-90}{\includegraphics[height=10cm]{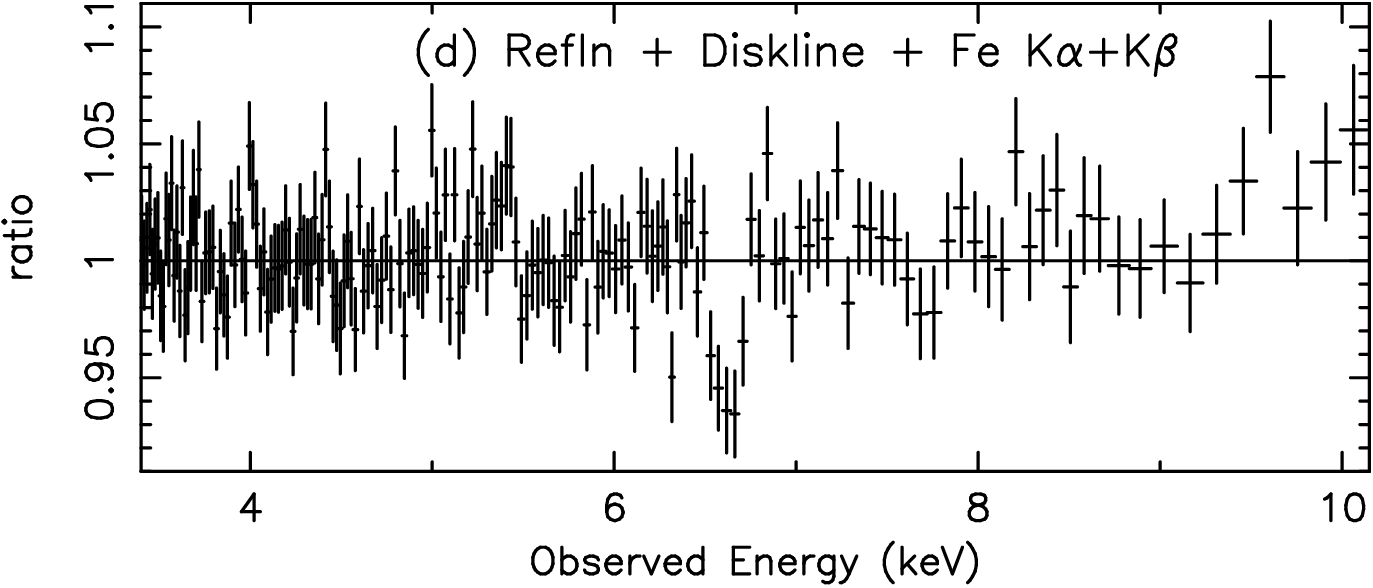}}
\end{center}
\caption{Data/model ratio residuals of the 3.5-12.0 keV band spectrum of 
NGC 3783, in the iron K-shell band. Panel (a) 
shows the residuals to a simple power-law fit, with Galactic absorption. 
The two strong peaks due to emission lines at 6.4 and 7.0 keV are readily 
apparent. Panel (b) shows the residuals after the two Fe emission lines 
at 6.4 keV and 7.0 keV have been fitted. 
Panel (c) shows the residuals after the addition of a neutral reflection
component with R=1, a weak red-wing to the Fe line is apparent below 6.4 keV. 
Panel (d) shows the residuals 
after the addition of a disk emission line component to model the red-wing. 
A strong absorption line, probably due to Fe~\textsc{xxv}, 
is present at 6.7 keV.}
\end{figure*}

\begin{figure*}
\rotatebox{-90}{ 
\epsscale{0.7}
\plotone{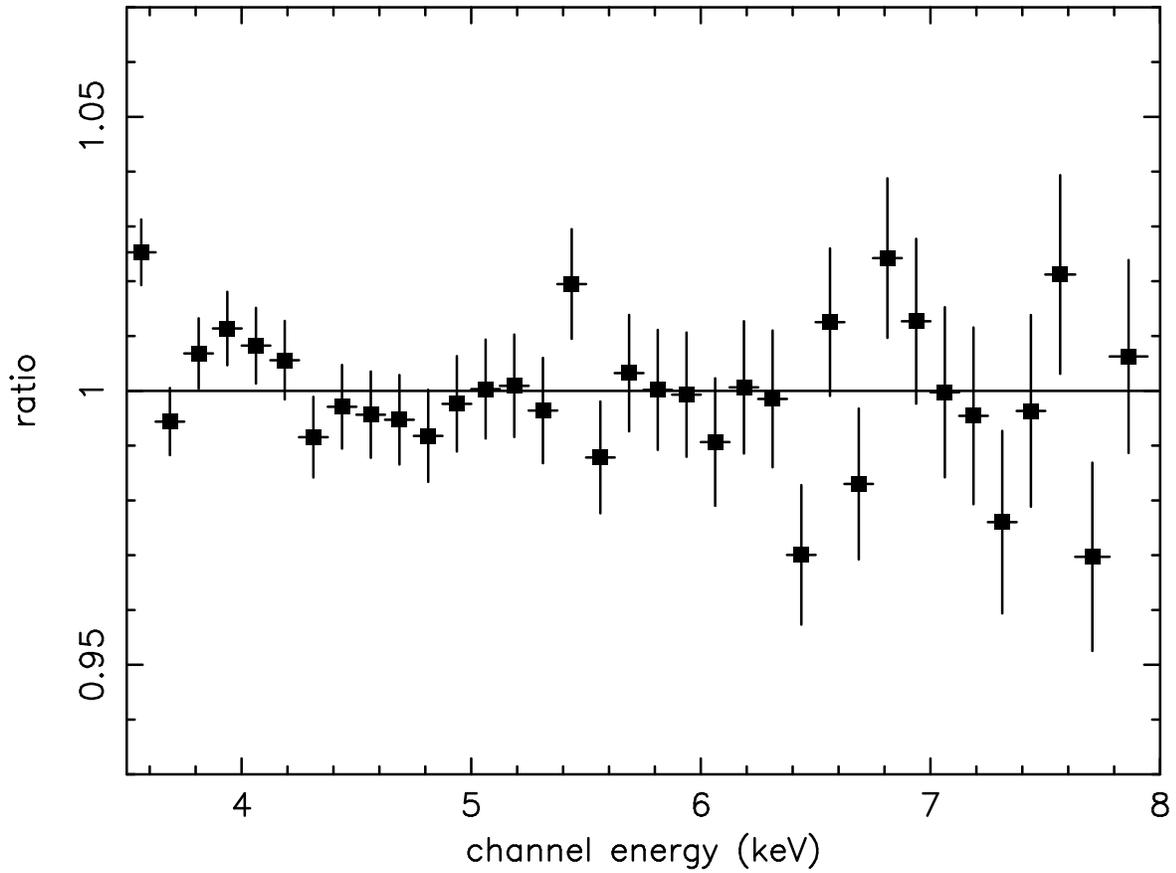}}
\caption{Data/Model ratio residuals to a power-law fit to a EPIC-pn 
small window calibration observation of the BL-Lac object, Mrk 421. 
No strong residuals are present over iron K bandpass to this featureless 
continuum source, indicating that the systematic residuals due to 
calibration are no more than 2\% of the continuum, over this energy range.}
\end{figure*}

\begin{figure*}
\rotatebox{-90}{ 
\epsscale{0.7}
\plotone{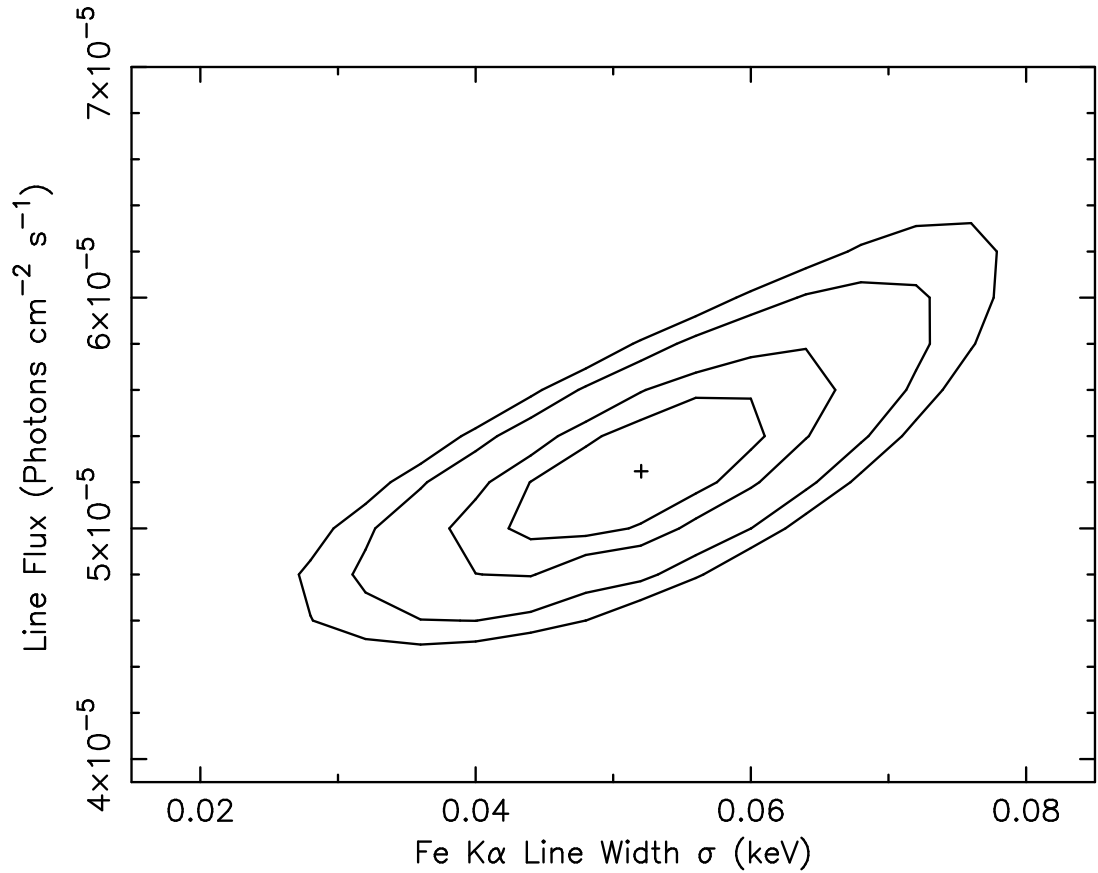}}
\caption{Contour plot showing the 68\%, 90\%, 99\% and 99.9\% 
confidence levels of Fe K$\alpha$ line flux versus line width for the 6.4 keV 
emission line. The line appears to be resolved by EPIC-pn with a width of 
$52\pm10$~eV, corresponding to a FWHM velocity width of 
$\sim5600$~km~s$^{-1}$. This places the bulk of the Fe line emission 
at the broad line region or even further away from the nucleus.}
\end{figure*}

\begin{figure*}
\rotatebox{-90}{
\epsscale{0.7}
\plotone{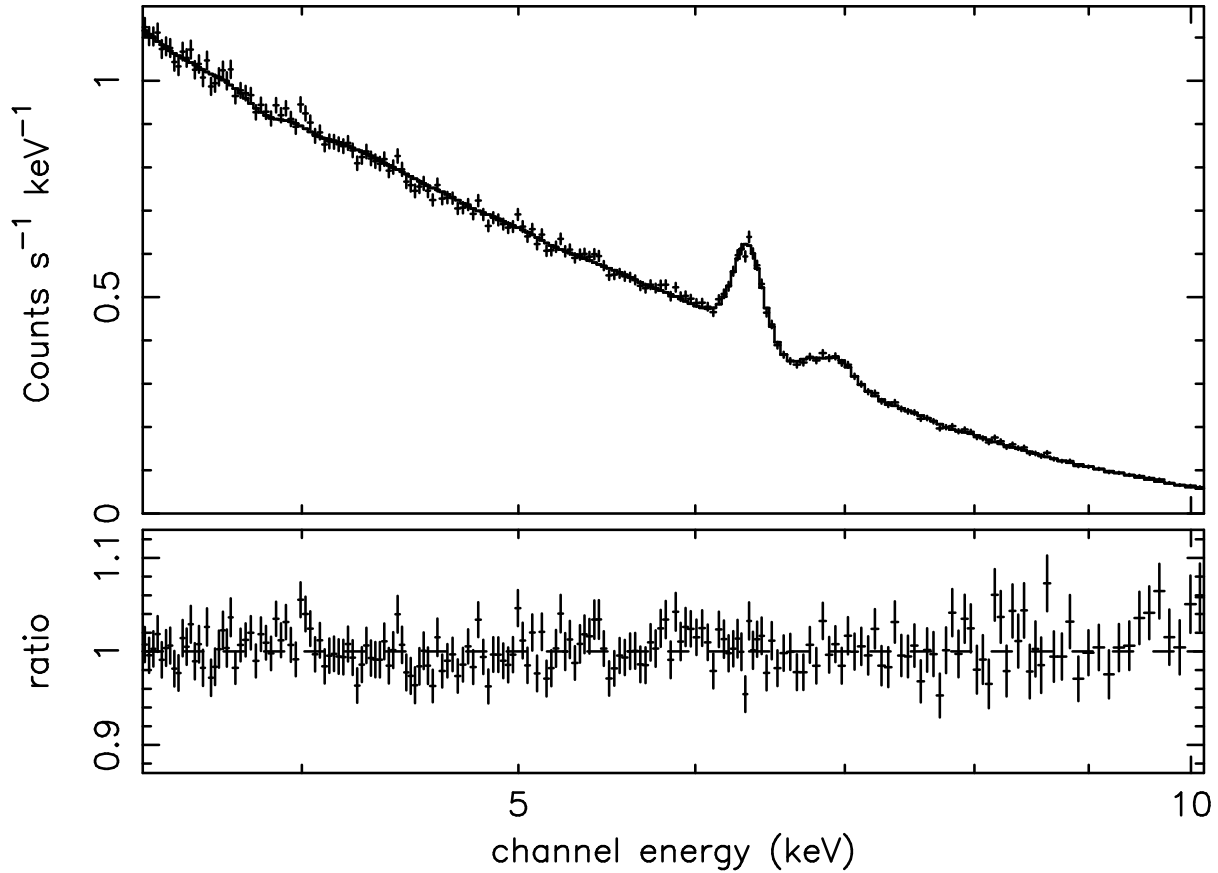}}
\caption{The 3.5-12 keV EPIC-pn spectrum (upper panel) and 
data/model residuals (lower panel)  
to a fit with a two zone warm absorber, modeled by the 
\textsc{xstar} photoionization code. The two narrow emission lines at 6.4 keV 
and 7.0 keV have been included in the fit, but the broad red-wing to the 
iron K$\alpha$ emission line is no longer required once the absorber has been 
correctly modeled. 
A high ionization absorber, with ionization parameter 
${\rm log}~\xi=2.9\pm0.3$~erg~cm~s$^{-1}$ and a column density 
of $N_{\rm H}=(4.6\pm0.8)\times10^{22}$~cm$^{-2}$ reproduces the 
energy and depth of the 6.7 keV absorption line rather well.}
\end{figure*}

\begin{figure*}
\rotatebox{-90}{
\epsscale{0.7}
\plotone{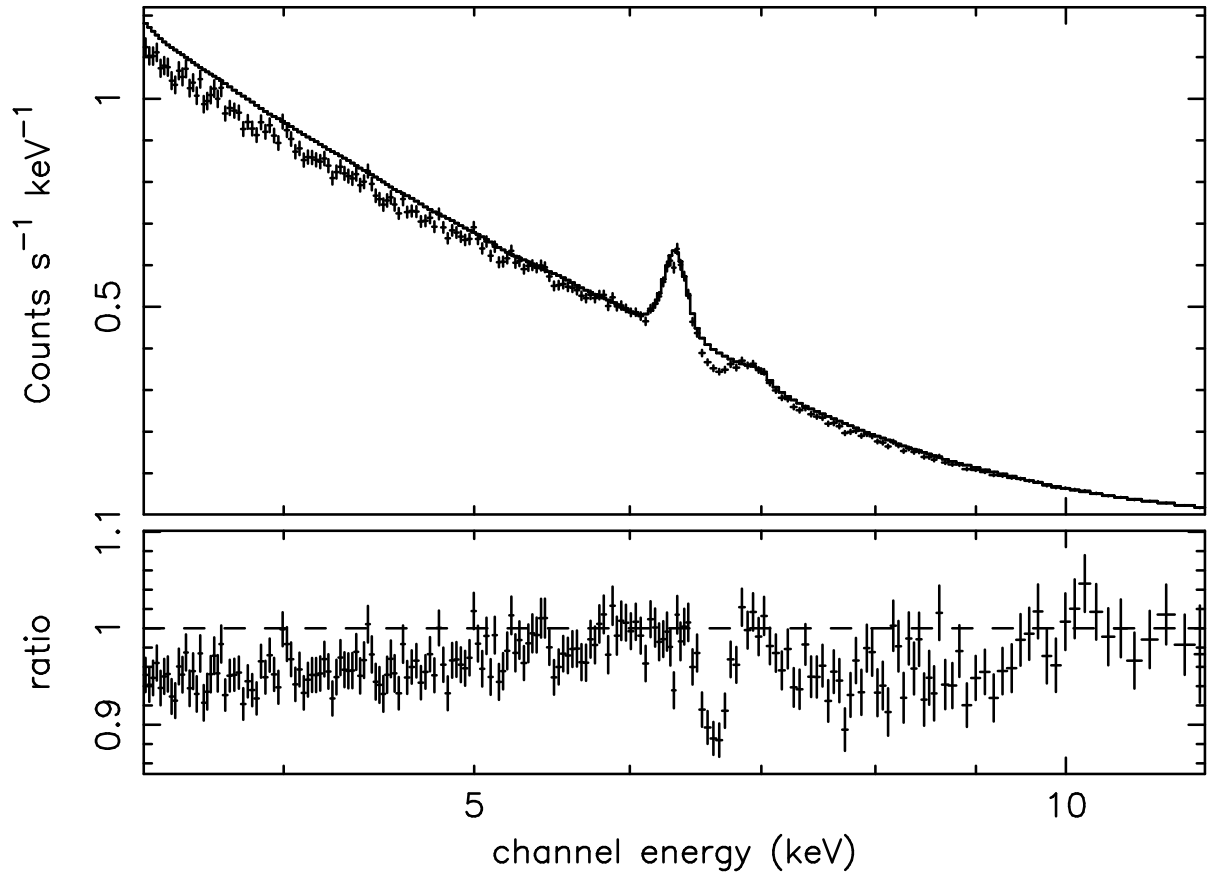}}
\caption{The 3.5-12 keV EPIC-pn spectrum and residuals after removing the 
two zone warm absorber in Figure 6, plotted in order to illustrate the effect 
of the ionized absorber on the hard X-ray spectrum. 
The deep absorption line at 6.7 keV is clearly apparent in the residuals, 
whilst an broad dip is present in the between 7 and 9 keV, 
due to iron K-shell edge absorption. 
There is also significant opacity in the absorber below 
6 keV, which effects any modeling of the broad iron line.}
\end{figure*}

\begin{figure*}
\rotatebox{-90}{
\epsscale{0.7}
\plotone{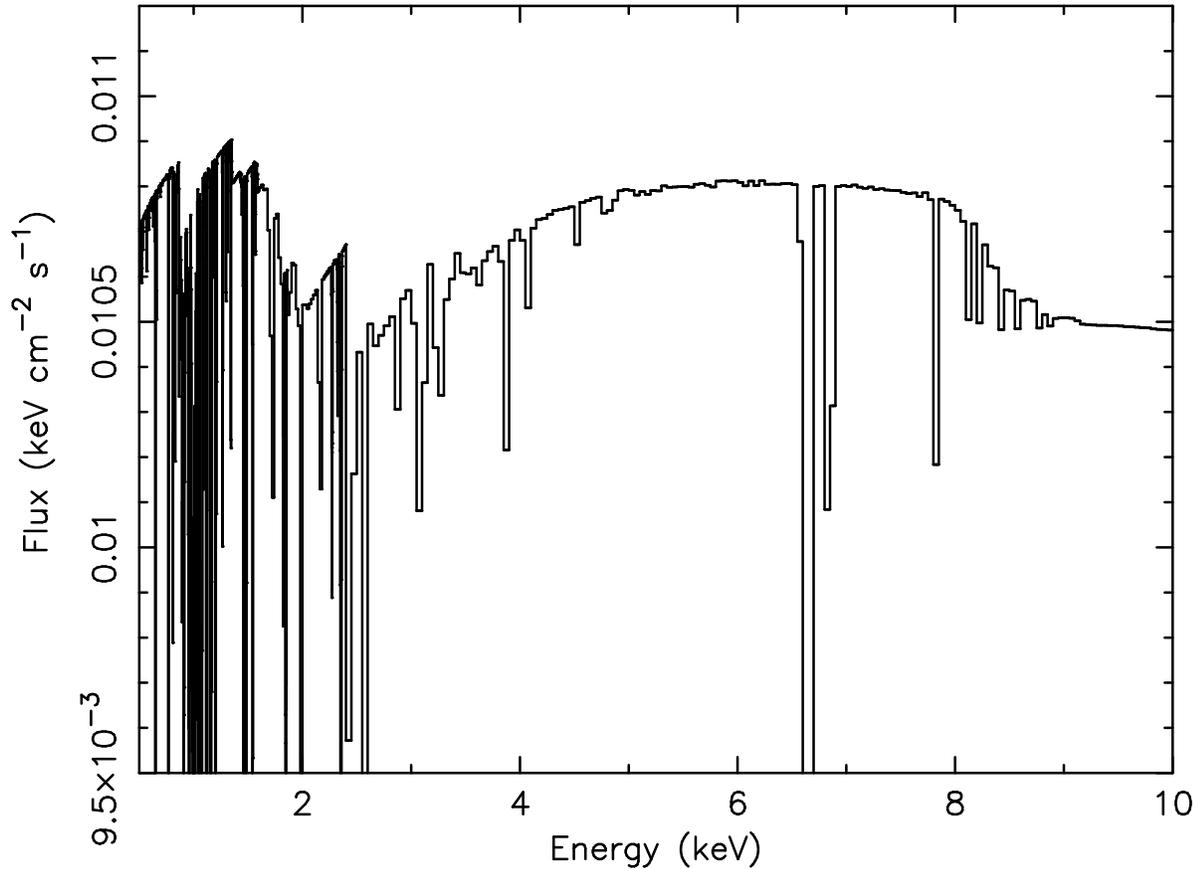}}
\caption{The high ionization warm absorber model, fitted to the 3.5-12 keV 
EPIC-pn spectrum, but normalized to a $\Gamma=2$ power-law continuum 
for ease of illustration only. 
As well as the strong Fe K$\alpha$ absorption lines that are 
present near 6.7 keV, 
the model introduces significant continuum curvature between 
3-6 keV, due to recovery from high ionisation K-shell edges and lines from 
Mg, Si, S, as well as L-shell Fe below 3 keV. Additional opacity is 
also between 7-9 keV, due to a blend of high ionization Fe K-shell edges and 
higher series absorption lines, such as Fe~K$\beta$.}
\end{figure*}

\begin{figure*}
\rotatebox{-90}{
\epsscale{0.7}
\plotone{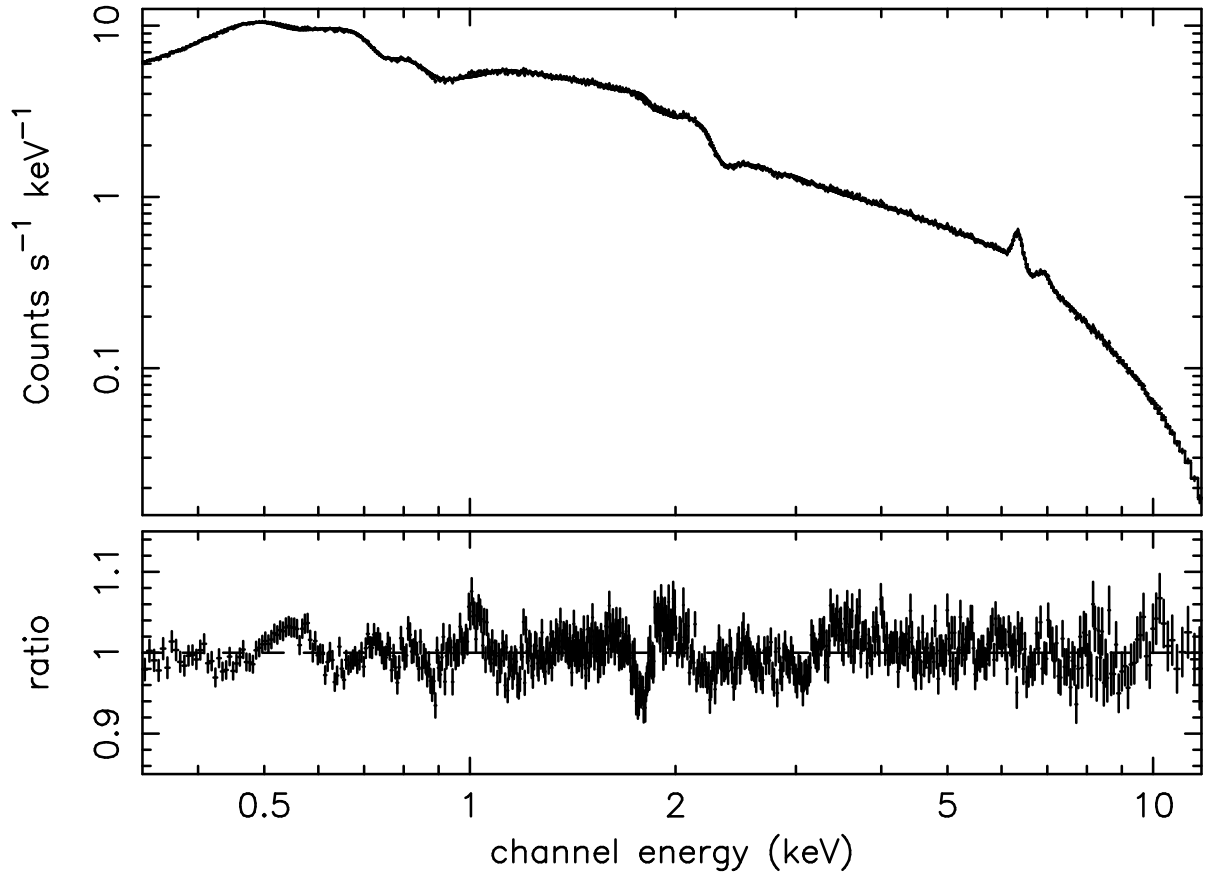}}
\caption{The broad-band 0.3-12 keV EPIC-pn spectrum and data/model 
residuals (lower panel). The spectrum has been modeled with a three  
component warm absorber. Two lower ionization components are required to 
model the  soft X-ray spectrum (with best-fit ionization parameters of 
log~$\xi\sim-0.1$ and log~$\xi\sim2.1$), 
as well as the high ionization component (with log~$\xi\sim3.0$) 
responsible for the Fe K-shell absorption. This 
model fits the 0.3-12~keV X-ray spectrum of NGC~3783 well, the remaining 
data/model residuals are at the 5\% level or smaller.}
\end{figure*}

\begin{figure*}
\rotatebox{-90}{
\epsscale{0.7}
\plotone{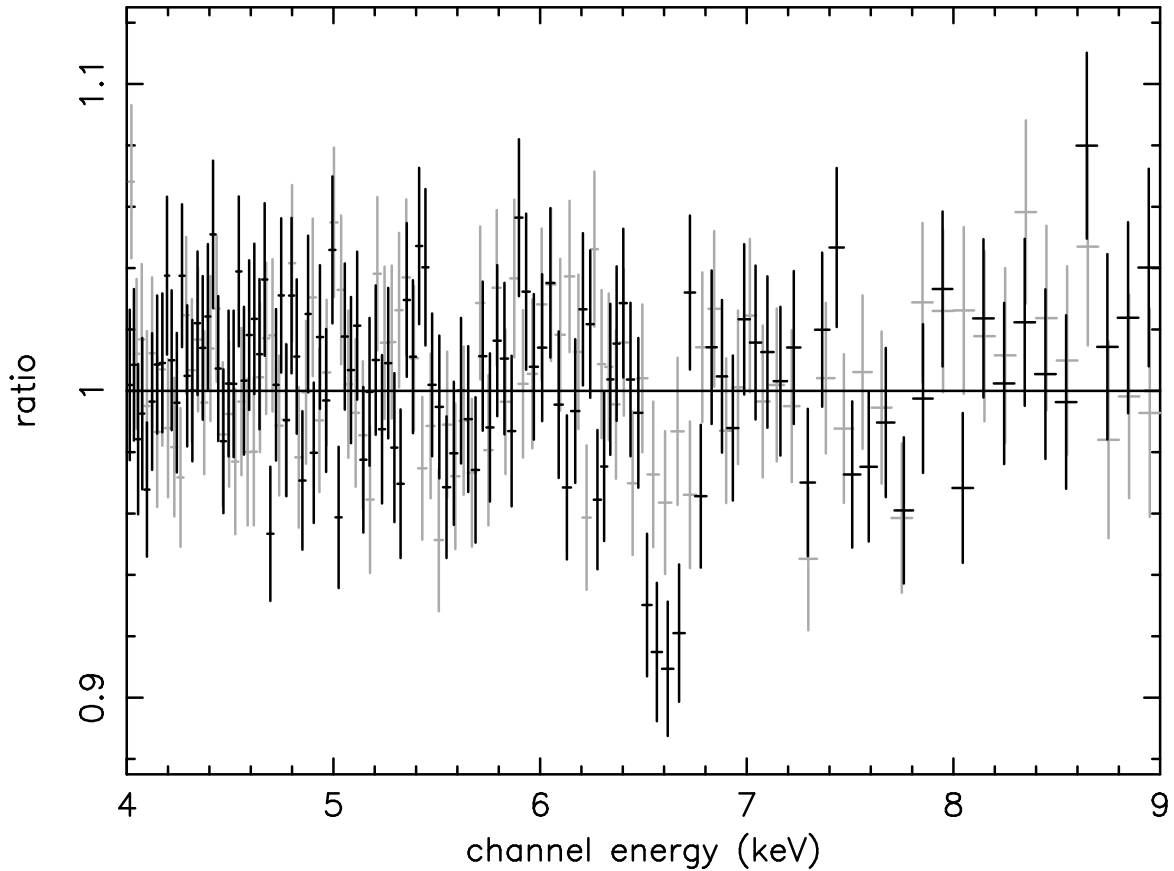}}
\caption{The data/model ratio residuals over the iron K-shell band from the 
two orbits of \xmm\ EPIC-pn data. The first orbit is plotted in greyscale, 
the second orbit in black. The narrow 6.4 and 7.0 keV emission lines 
have been modeled in the spectrum. The 6.7 keV absorption line is clearly 
present in the second orbit of data, but is very weak in the first orbit, 
indicating that this high ionization absorption component is 
variable on short timescales, of the order $10^{5}$~s.}
\end{figure*}

\begin{figure*}
\rotatebox{-90}{
\epsscale{0.75}
\plotone{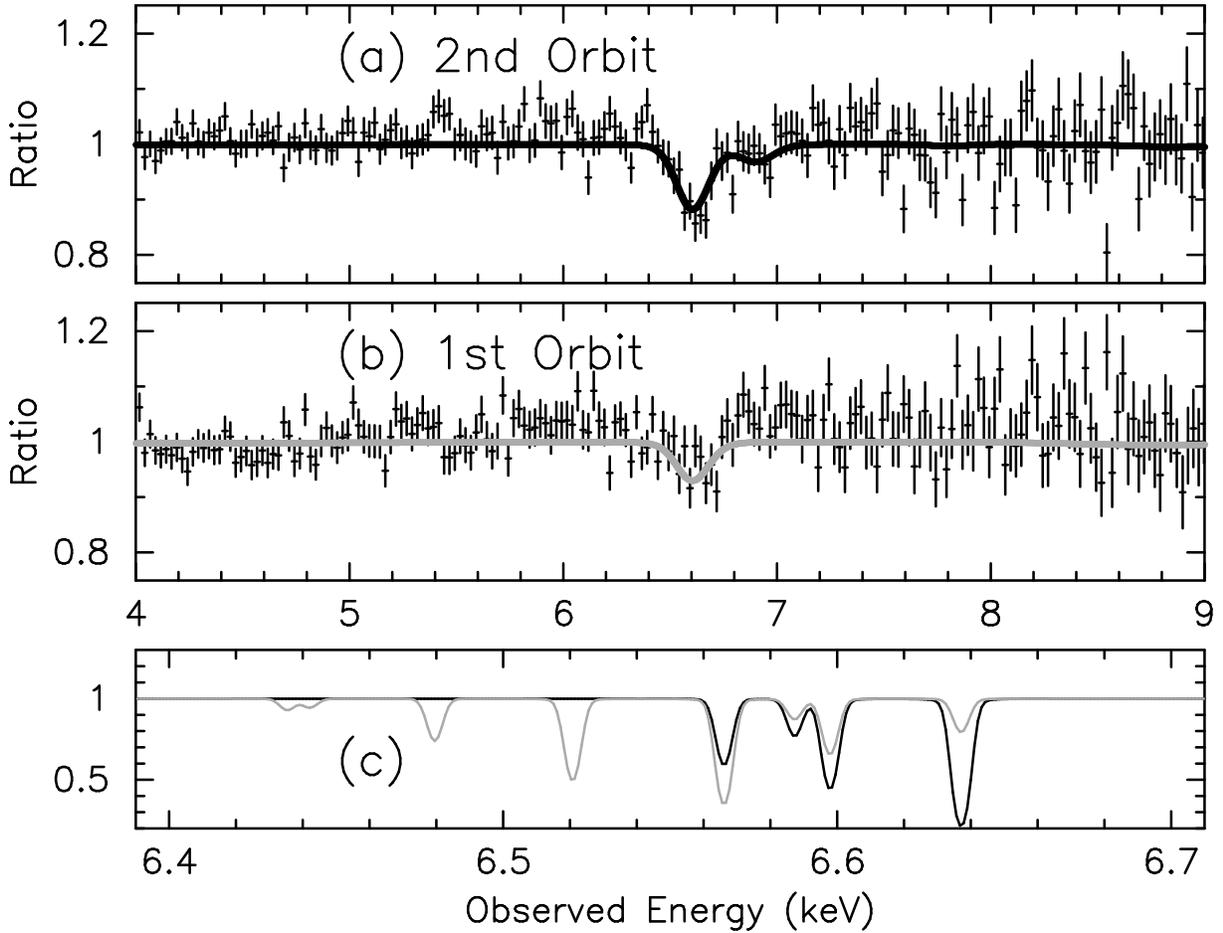}}
\caption{Fits to the iron K-shell absorption line in the 
two seperate \xmm\ orbits.  
(a) The data (crosses) and absorption line model (solid line), convolved 
through the EPIC-pn response,  
to the \xmm\ spectrum during the 2nd orbit. Here, the
absorption line is strongest when the continuum flux is higher. 
It can be modeled with a high ionization 
absorber, where the Fe~\textsc{xxv} resonance ($1s \rightarrow 2p$) 
line is the strongest line present. 
(b) The first orbit of data, when the continuum flux is lower 
and best fit model overlaid (solid grey line). The absorption line 
is weaker during the 1st orbit, which is consistent with a lower 
overall ionization for the iron K band absorber. 
(c) The relative strengths of the absorption 
lines from Fe~\textsc{xxv} to Fe~\textsc{xx} (right to left) from the 
two models shown above; the lines correspond to observed energies of 6.64 keV, 
6.59 keV (doublet), 6.565 keV, 6.52 keV, 6.48 keV and 6.44 keV, 
or rest (lab) frame energies of 6.70 keV, 6.66 keV, 
6.63 keV, 6.58 keV, 6.54 keV and 6.50 keV respectively. The high ionization 
model (for the 2nd orbit) is plotted in black, the lower ionization model 
(for the 1st orbit) in greyscale. Thus the higher flux 
2nd orbit may be 
dominated by the higher ionization ions (Fe~\textsc{xxiii--xxv}), whilst 
the lower ionization species (Fe~\textsc{xx--xxiii})
are more dominant in the lower flux 1st orbit spectrum.
}
\end{figure*}

\clearpage

\begin{deluxetable}{lccccccc}
\tabletypesize{\small}
\tablecaption{Table of iron line spectral fits.}
\tablewidth{0pt}
\tablehead{
\colhead{Fit} & \colhead{$\Gamma$} & \colhead{Line} &
\colhead{$E$\tablenotemark{a}} & \colhead{$\sigma$\tablenotemark{b}} & 
\colhead{$EW$\tablenotemark{c}} &
\colhead{$\chi^{2}/dof$}  &
\colhead{$P_{\null}$\tablenotemark{d}} }

\startdata

1. & $1.62\pm0.01$ & Fe~K$\alpha$ & 
$6.39\pm0.01$ & $57\pm8$ & $123\pm6$ & 
1589/1287 & $1.4\times10^{-8}$ \\

& & Fe~K$\beta$/Fe~\textsc{xxvi} &  
$7.00\pm0.02$ & $53\pm23$ & $34\pm5$ \\ 

\\

2. & $1.62\pm0.01$ & Fe~K$\alpha$ & 
$6.39\pm0.01$ & $57\pm8$ & $123\pm6$ & 
1591/1287 & $1.15\times10^{-8}$ \\

& & Fe~\textsc{xxvi} & 
$6.96\pm0.02$ & $57\tablenotemark{e}$ & $20\pm5$ \\

& & Fe~K$\beta$ & 
$7.06\tablenotemark{e}$ & $57\tablenotemark{e}$ & $16\tablenotemark{e}$ \\

\\

3. & $1.73\pm0.01$ & Fe~K$\alpha$ & 
$6.39\pm0.01$ & $47\pm8$ & $109\pm5$ & 
1492/1287 & $5.6\times10^{-5}$ \\

& & Fe~\textsc{xxvi} & 
$7.00\pm0.04$ & $47\tablenotemark{e}$ & $12\pm4$ \\

& & Fe~K$\beta$ & 
$7.06\tablenotemark{e}$ & $47\tablenotemark{e}$ & $14\tablenotemark{e}$ \\

\\

4. & $1.73\pm0.01$ & Fe~K$\alpha$ & 
$6.40\pm0.01$ & $53\pm8$ & $107\pm8$ & 
1324/1282 & $2.0\times10^{-1}$ \\

& & Fe~\textsc{xxvi} & 
$6.98\pm0.04$ & $53\tablenotemark{e}$ & $17\pm5$ \\

& & Fe~K$\beta$ & 
$7.06\tablenotemark{e}$ & $53\tablenotemark{e}$ & $14\tablenotemark{e}$ \\

& & Fe~\textsc{xxv} absn & 
$6.67\pm0.04$ & $10\tablenotemark{e}$ & $17\pm5$ \\

& & Diskline & 
$6.4\tablenotemark{e}$ & & $58\pm12$ \\

\enddata


\tablenotetext{a}{Energy of the Fe line in units of keV.} 
\tablenotetext{b}{$1\sigma$ width of the line in eV.}
\tablenotetext{c}{Equivalent width of the line in eV.}
\tablenotetext{d}{Null hypothesis probability that the fit statistic 
is acceptable.}
\tablenotetext{e}{Parameter value is fixed, or tied to that of another fit 
parameter} 

\tablecomments{Fit 1 consists of a power-law plus 2 
Gaussian emission lines. Fit 2 is a power-law plus 3 Gaussian 
emission lines. Fit 3 is a power-law plus 3 Gaussian emission lines and 
neutral reflection component (the \textsc{xspec} model 
\textsc{pexrav} with $R=1$). Fit 4 consists of a power-law, plus three 
Gaussian emission lines, a neutral reflection component, a disk emission line 
(the \textsc{diskline} model in xspec), as well as a Gaussian shaped 
absorption line.}

\end{deluxetable}

\clearpage

\end{document}